\documentclass[pdftex]{pasj01}
\draft 

\Received{$\langle$reception date$\rangle$}
\Accepted{$\langle$acception date$\rangle$}
\Published{$\langle$publication date$\rangle$}

\usepackage{natbib}

\usepackage{aas_macros}
\usepackage{lineno}

\usepackage{url}

\begin{document}

\title{Probing the primordial Universe with 21-cm line from cosmic dawn/epoch of reionization}

\author{Teppei \textsc{Minoda}\altaffilmark{1}\altaffilmark{2}}%
\author{Shohei \textsc{Saga}\altaffilmark{3}}%
\author{Tomo \textsc{Takahashi}\altaffilmark{4}}%
\author{Hiroyuki \textsc{Tashiro}\altaffilmark{2}}%
\author{Daisuke \textsc{Yamauchi}\altaffilmark{5}}%
\author{Shuichiro \textsc{Yokoyama}\altaffilmark{6}\altaffilmark{7}}%
\author{Shintaro \textsc{Yoshiura}\altaffilmark{8}\altaffilmark{1}}%

\email{minoda@nagoya-u.jp}
\altaffiltext{1}{The University of Melbourne, School of Physics, Parkville, VIC 3010, Australia}
\altaffiltext{2}{Department of Physics and Astrophysics,
Nagoya University, Chikusa-ku, Nagoya, 464-8602, Japan}
\altaffiltext{3}{Laboratoire Univers et Th{\'e}ories, Observatoire de Paris, Universit{\'e} PSL, Universit{\'e} de Paris, CNRS, F-92190 Meudon, France}
\altaffiltext{4}{Department of Physics,Saga University, Saga 840-8502,Japan}
\altaffiltext{5}{Faculty of Engineering, Kanagawa University, Kanagawa, 221-8686, Japan}
\altaffiltext{6}{Kobayashi Maskawa Institute, Nagoya University, Aichi 464-8602, Japan}
\altaffiltext{7}{Kavli Institute for the Physics and Mathematics of the Universe (WPI),
Todai institute for Advanced Study, University of Tokyo, Kashiwa, Chiba 277-8568, Japan}
\altaffiltext{8}{Mizusawa VLBI Observatory, National Astronomical Observatory Japan,
2-21-1 Osawa, Mitaka, Tokyo 181-8588, Japan}


\maketitle

\begin{abstract}
In the most distant reaches of the Universe, the 21-cm hyperfine transition in neutral hydrogen provides
one of the only available tracers of large-scale structure.
A number of instruments have been working and planned to measure the 21-cm line signals, and
in particular, Experiment to Detect the Global EoR Signature (EDGES) recently has reported the first detection
of an absorption signal, which corresponds to the 21-cm line global signal at the epoch of reionization (EoR).
The future large radio telescope, Square Kilometre Array (SKA) will be able to deliver 
the high-precision measurement of 21-cm line emission/absorption signals.
In this paper, we review the current status for the 21-cm line global and fluctuation signals from EoR to the dark ages, and then summarize the recent studies of how we probe the primordial Universe particularly motivated by the recent EDGES result and future observations by SKA.
We focus on two applications for constraining cosmology with the EDGES result: constraints on the primordial 
magnetic fields and those on the primordial power spectrum. 
We also discuss the potential of future SKA for probing the inflationary Universe, 
by discussing expected constraints on the primordial power spectrum, its adiabaticity, and primordial non-Gaussianities 
from future observations of 21-cm fluctuations. 
\end{abstract}


\section{Introduction}

Inflation, the accelerated expansion phase of the early universe, has been widely studied as a standard paradigm that can naturally address the shortcomings of Big Bang cosmology.
In particular, recent observations of anisotropies of cosmic microwave background (CMB) radiation, cosmic large-scale structure, and so on strongly support the inflationary mechanism as the origin of primordial density fluctuations.
In fact, by scrutinizing the observational data, we have been able to impose constraints on a bunch of inflation models (see, e.g., \cite{Martin:2013tda}).
However, it is still not enough to fully understand the inflation mechanism, and more information needs to be extracted from observations.

In this situation, high-precision deep-Universe exploration by large radio telescopes, such as the future Square Kilometre Array (SKA)~\footnote{\url{https://www.skatelescope.org}}, is of great importance. The SKA consists of two types of telescopes; SKA-LOW and SKA-MID. A primary information of the SKA can be found in SKA Phase 1 Construction Proposal \footnote{\url{https://www.skatelescope.org/wp-content/uploads/2021/02/22380_Construction-Proposal_DIGITAL_v3.pdf}}. 
The SKA-LOW will be built at Murchison Radio-astronomy Observatory (MRO) in Western Australia. The SKA-LOW will consist of 512 stations and each station arrange 256 log-periodic antennas (i.e. 131,072 antennas in total). The SKA-MID, on the other hand, will be built in the Northern Cape of South Africa. The SKA-MID will consist of 133 SKA 15-m dishes and 64 MeerKAT 13.5-m dishes. The observable frequency range is from 50~MHz to 350~MHz for the SKA-LOW and from 350~MHz to 15.3~GHz for the SKA-MID. The sensitivity of SKA telescopes will be roughly 10 times better than current telescopes. With the excellent sensitivity and the wide field of view, the survey speed is also more than 10 times better than current telescopes. The prospects of SKA to probe various aspects of cosmology have been summarized in e.g., \cite{Yamauchi:2016ypt,Bacon:2018dui}.

Among various observational modes which are delivered by SKA, 
observations of the redshifted 21-cm line of neutral hydrogen (HI) open up a new window for observational cosmology. The SKA-LOW will measure the 21-cm line from intergalactic medium (IGM) during cosmic dawn and epoch of reionization (EoR). The SKA-MID with band 1 (350~MHz--950~MHz) will measure the 21-cm line associated with galaxies at $0.5<z<3$ using the technique of intensity mapping to measure the large-scale HI distribution rather than resolving individual galaxies.
The SKA is expected to measure
the spatially fluctuating 21-cm line signal with a very high resolution,
and achieve a lot of scientific goals
(e.g., measuring the equation-of-state of dark energy,
constraining modified theories of gravity,
anisotropy and inhomogeneity of the universe,
and statistical properties of the matter density fluctuations,  
measuring the HI density and bias, and so on;
for more detailed scientific goals with SKA,
we refer the readers to \citealt{2020PASA...37....7S}). 
A number of instruments have been working and planned to measure the 21-cm line fluctuation at post reionization era such as the Baryon Acoustic Oscillations in Neutral Gas Observations (BINGO, \cite{2013MNRAS.434.1239B}), the Hydrogen Intensity and Real-Time Analysis experiment (HIRAX, \cite{2016SPIE.9906E..5XN}), Five-hundred-meter Aperture Spherical Radio Telescope (FAST, \cite{2011IJMPD..20..989N,2020MNRAS.493.5854H}), Canadian Hydrogen Intensity Mapping Experiment (CHIME, \cite{2014SPIE.9145E..22B}), Australian SKA Pathfinder (ASKAP, \cite{2012MNRAS.426.3385D}), MeerKAT \citep{2017arXiv170906099S}, and to detect the signal before the reionization such as the Precision Array for Probing the Epoch of Reionization (PAPER, \cite{Parsons2010TheResults2}), the Giant Metrewave Radio Telescope EoR Experiment (GMRT, \cite{Paciga2013AExperiment2}), the Murchison Widefield Array (MWA, \cite{Tingay2013TheFrequencies,Wayth2018TheOverview2}), the LOw Frequency ARray (LOFAR, \cite{VanHaarlem2013LOFAR:Array2}), and the Hydrogen Epoch of Reionization Array (HERA, \cite{Deboer2017HydrogenHERA2}). 
Not only the fluctuations of the 21-cm line signal,
but its intensity averaged over all-sky,
which is the so-called 21-cm global signal,
also provides 
important cosmological information
\citep{1999A&A...345..380S,2006MNRAS.371..867F,2010PhRvD..82b3006P,2015ApJ...813...11M,2016MNRAS.457.1864L,2017MNRAS.472.1915C}.
The measurement of the 21-cm global signal can be achievable with a single dipole antenna. For the ground-based 21-cm global signal measurement, active developments of instruments are planned and working; Experiment to Detect the Global EoR Signature (EDGES, \cite{2017ApJ...847...64M}), 
Broadband Instrument for Global HydrOgen ReioNisation Signal (BIGHONS, \cite{2015PASA...32....4S}), 
Shaped Antenna measurement of the background RAdio Spectrum (SARAS, \cite{2013ExA....36..319P,2021arXiv210401756N}), 
Sonda Cosmol\'{o}gica de las Islas para la Detecci\'{o}n de Hidr\'{o}geno Neutro
 (SCI-HI, \cite{2014ApJ...782L...9V}), 
the Large-aperture Experiment to detect the Dark Ages (LEDA; \cite{2015ApJ...799...90B,2016MNRAS.461.2847B,2018MNRAS.478.4193P}), 
the Cosmic Twilight Polarimeter (CTP, \cite{2017ApJ...836...90N,2019ApJ...883..126N}),
Zero-spacing Interferometer Measurements of the Background Radio Spectrum (ZEBRA; \cite{2014arXiv1406.2585M}), 
Radio Experiment for the Analysis of Cosmic Hydrogen (REACH, \cite{8879199}), 
Mapper of the IGM Spin Temperature (MIST, \url{http://www.physics.mcgill.ca/mist/}), and
Probing Radio Intensity at high z from Marion (PRI$^{\rm Z}$M, \cite{2019JAI.....850004P}). For the EDGES Low band observation, a blade-shaped antenna is deployed located in the MRO in Western Australia. 
Recently, \cite{2018Natur.555...67B} has reported
the first detection of
an absorption profile around 78~MHz with EDGES,
which corresponds to the 21-cm signal from $z\simeq 17$.
This is the first report of the 21-cm line signal before EoR.

In this review paper, we focus on how current and future observations of 
21-cm line emission/absorption signals of neutral hydrogen 
from cosmic dawn/EoR can elucidate the primordial Universe. Particularly 
we discuss how the 21-cm observation with large radio telescopes such as SKA will contribute to the understanding of the inflation mechanism. 
Indeed, we show that the nature of primordial fluctuations generated from inflation such as the power spectrum, 
the adiabaticity and non-Gaussianity can be well probed by future observations of SKA. 
We also discuss what kind of cosmological information we can obtain
and how we can extract it from the 21-cm global signal observations.
Since the 21-cm global signal is related to the IGM gas temperature,
the heating source for the IGM gas beyond the standard 
cosmology, so-called $\Lambda$-cold dark matter (CDM),
can be strongly constrained by observations of the 21-cm line.
In addition, Lyman-$\alpha$ radiation
from the first astrophysical object can also affect the 21-cm signal,
which depends on how the structure formation proceeds
and hence cosmological scenario changing density fluctuations
would also be constrained by the global signal.
In particular, the recent EDGES observation has stimulated
a lot of works along the line of the above consideration,
which we aim to overview in this paper.

This paper is organized as follows:
In section \ref{sec:Cosmological 21cm-line signals}, we briefly review 
the current status for the 21-cm global signals and 21-cm fluctuations.
We first describe how the 21-cm signals can be theoretically estimated in the standard cosmological models,
and we then summarize their cosmological implication particularly motivated by the recent EDGES result and
future observations such as SKA.
In section \ref{sec:inflation}, we consider the application to the primordial spectrum that encodes the physical
information about the inflationary Universe. In particular, we discuss 
how precisely the inflationary
model parameters can be constrained by SKA and
more futuristic observations such as Fast Fourier Transform Telescope (FFTT) or Omniscope \citep{Tegmark:2008au,Tegmark:2009kv}.
In section \ref{sec:PMFs}, we focus on the primordial magnetic fields (PMFs) as another key to understand
the primordial Universe.
We review the effect of PMFs on the thermal evolution and dynamics of the IGM gas at EoR and dark ages, and discuss the constraint
on PMFs from future 21-cm observations.
The final section is devoted to the summary. 

\section{Cosmological 21-cm line signals}
\label{sec:Cosmological 21cm-line signals}

This section describes the basics of the 21-cm global signal and fluctuations. There are three scopes: 21-cm global IGM signal,
21-cm fluctuations from inhomogeneous IGM, and those from galaxies.
These three observables are observed with different types of telescopes and frequencies. 

At first, we introduce the general form of the 21-cm signal produced from neutral hydrogen atoms in the IGM with cosmic microwave background as a backlight.
This is conventionally described by the differential brightness temperature relative to the CMB which can be given by (e.g. \cite{Furlanetto:2006jb}),
\begin{equation}
\delta T_{b{\rm (IGM)} } 
(\boldsymbol{x})
= \frac{3 hc^3 A_{10} x_{\rm HI}(\boldsymbol{x}) n_{\rm H} (\boldsymbol{x})}{32 \pi k_{\rm B} \nu_{21}^2 (1+z)^2 ({\rm d} v_\parallel /{\rm d}r)}
\left ( 1- \frac{T_{\rm CMB}}{T_{\rm s} (\boldsymbol{x})} \right) \,,
\label{eq:IGMbrightness}
\end{equation}
where $h$ is the Planck constant, $c$ is the speed of light, $A_{10}$ is the spontaneous decay rate of 21-cm transition, $x_{\rm HI}(\boldsymbol{x})$ is the neutral fraction of hydrogen atoms, $n_{\rm H}(\boldsymbol{x})$ is the number density of hydrogen, $k_{\rm B}$ is the Boltzmann constant, $\nu_{21}$ is the frequency corresponding to 21-cm line, ${\rm d}v_\parallel / {\rm d}r $ is the velocity gradient along the line of sight,  $T_{\rm CMB}$ and $T_{\rm s}(\boldsymbol{x})$ are the CMB temperature and spin temperature, respectively. The spin temperature is defined by
the ratio of the population of the hyperfine levels in neutral hydrogen as
\begin{equation}
\frac{n_1}{n_0} = \frac{g_1}{g_0}
\exp \left(-\frac{\Delta T_{10}}{T_{\rm s}}\right)~.
\end{equation}
Here, $n_1$ and $n_0$ are number densities
in 1S singlet and 1S triplet states,
and $g_1$ and $g_0$ are their statistical weights, respectively. $\Delta T_{10}= 0.068~{\rm K}$ is the temperature corresponding to the energy of 21-cm photon.

\subsection{21-cm global signal}

The observable of the 21-cm line global signal
is the all-sky averaged differential brightness temperature.
By averaging equation (\ref{eq:IGMbrightness}) over the spherical surface at redshift $z$,
we obtain
\begin{eqnarray}
\overline{\delta T}_{\rm b{\rm (IGM)}} &(z)
&\simeq  27 \bar{x}_{\rm HI}(z) 
\left[1 - \frac{T_{\rm CMB}(z)}{T_{\rm s}(z)}\right]
\nonumber \\
&\times& \left(\frac{\Omega_{\rm b} h^2}{0.02}\right)
\left(\frac{0.15}{\Omega_{\rm m} h^2}\right)^{1/2}
\left(\frac{1+z}{10}\right)^{1/2}\, [{\rm mK}]~,
\label{dTb}
\end{eqnarray}
where $\bar{x}_{\rm HI}(z)$ is the averaged neutral fraction of hydrogen atoms.

When the spin temperature is equal to the CMB temperature,
the 21-cm signal disappears as shown in equation~(\ref{dTb}). 
In the cosmological context,
the spin temperature deviates from the CMB temperature via two processes such as the collisional interaction and the interaction with Lyman-$\alpha$ (Ly-$\alpha$) radiation \citep{1952AJ.....57R..31W,1959ApJ...129..536F}.
As these processes can couple the spin temperature with the gas temperature in the IGM, the spin temperature evolves between the values of the CMB and the IGM gas temperatures which are given as
\citep{2012RPPh...75h6901P}
\begin{equation}
T_{\rm s}^{-1}=\frac{T_{\rm CMB}^{-1}+x_{\alpha}
T_{\alpha}^{-1}+x_{c} T_{\rm K}^{-1}}{1+x_{\alpha}+x_{c}}~.
\label{T_spin}
\end{equation}
Here, $x_{\alpha}$ and $x_c$ are coupling coefficients for Ly-$\alpha$ interaction and collisional interaction, respectively. $T_\alpha$ is the color temperature of the background Ly-$\alpha$ photons. Due to the repeated resonant scattering of Ly-$\alpha$ photons with the hydrogen atoms, $T_\alpha$ is expected to be coupled with the kinetic gas temperature $T_{\rm K}$. Indeed other effects might contribute to the evolution history of the spin temperature.

\begin{figure}
\centering
\includegraphics[width=0.6\textwidth]{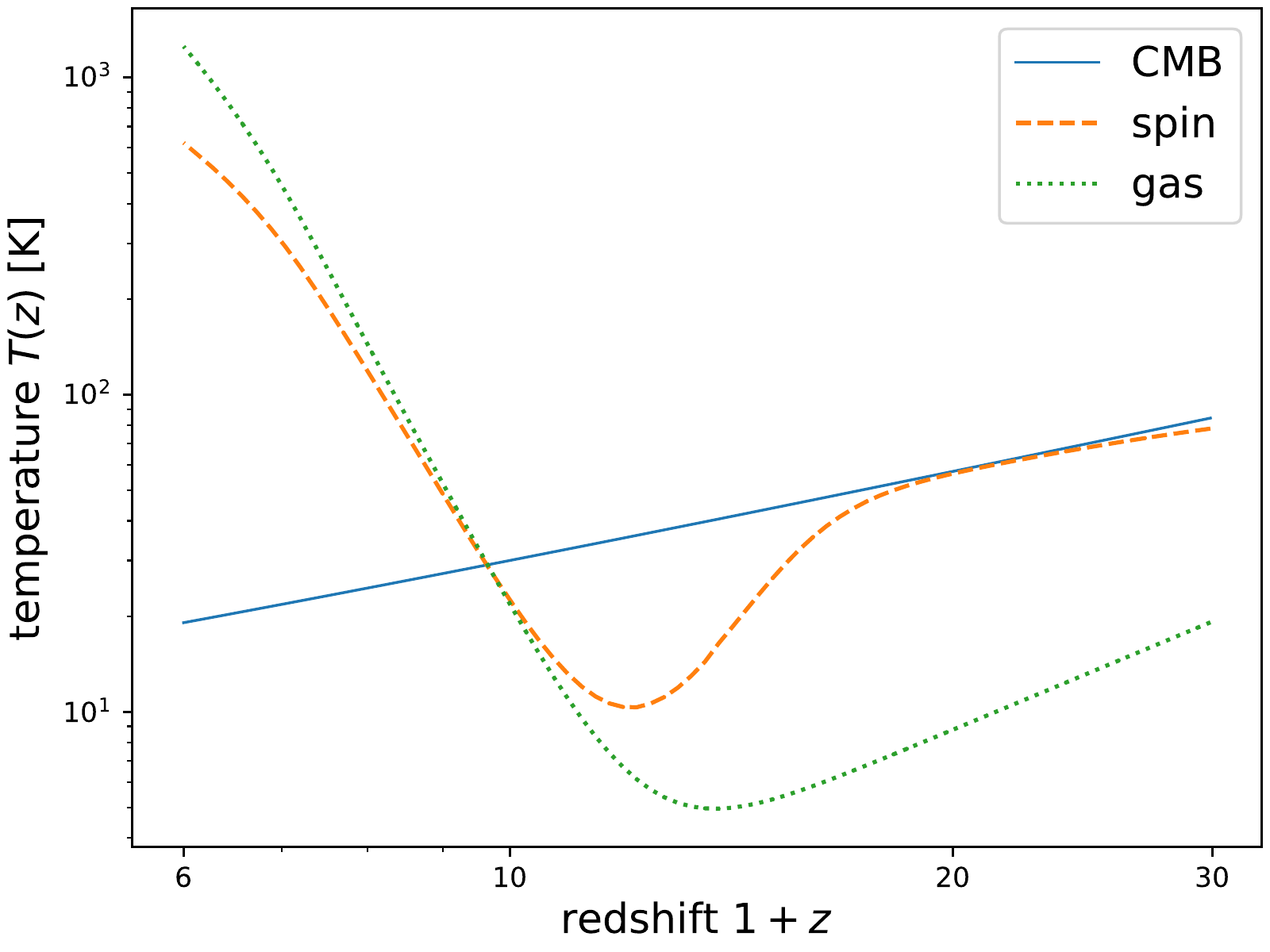}
\vspace{-0mm}
\caption{
Fiducial redshift evolution of the spin, CMB and gas temperature 
which are calculated using \texttt{21cmFAST}
\citep{2011MNRAS.411..955M,2020JOSS....5.2582M}.
The solid, dashed, and dotted lines are CMB, spin, and gas temperature, respectively.
The spin temperature starts to couple with the kinetic gas temperature due to the Ly-$\alpha$ coupling at $z\lesssim 20$. The 21-cm signal would be an absorption line during the spin temperature is less than the CMB one.
}
\label{fig:T_spin}
\end{figure}

Figure \ref{fig:T_spin} shows the evolution of
the spin temperature in a fiducial model, which has been calculated by using
{\tt {21cmFAST}}~\citep{2011MNRAS.411..955M}.
Here, we describe {\tt {21cmFAST}}
\footnote{We have used the version 2 of the {\tt {21cmFAST}}.The latest version \citep{2020JOSS....5.2582M} is available at {https://github.com/21cmfast/21cmFAST}}
briefly. For details about the calculation and assumption of the \texttt{21cmFAST}, the readers 
are referred to \cite{2019MNRAS.484..933P, 2020MNRAS.495..123Q}.
Following Zel'dovich approximation, matter density distributions are generated from an initial condition given in a high-resolution box.
Ionization is solved using the semi-numerical method
FFRT~\citep{2011MNRAS.414..727Z}
which is based on the excursion-set approach 
\citep{2004ApJ...613....1F}. Specifically, the number of ionizing photons, neutral hydrogen atoms, and the cumulative number of recombination are evaluated within a region with a radius. If the number of ionizing photons is enough to ionize the region, the centre of the region is flagged as ionized. By decreasing the radius, {\tt{21cmFAST}} repeats this procedure. The evolution of spin temperature is also derived
by solving the ionized fraction and kinetic gas temperature in the neutral regions with the evaluation of the Ly-$\alpha$, ionization rate, and X-ray backgrounds \citep[see][]{2011MNRAS.411..955M}. 
To evaluate the total amount of UV and X-ray emissions, 
the Press-Schechter halo mass function is employed. The emission from halo is modeled with parameters related to the astrophysical model such as the fraction of gases converted to stars $f_*$, and power-law index of $f_*$, escape fraction of ionizing photons $f_{\rm esc}$,
power-law index of $f_{\rm esc}$, time scale of star formation relative to the Hubble time, X-ray luminosity per star formation rate,  minimum energy of X-ray emission and turn over halo mass scale $M_{\rm turn}$. The emission from halo with masses below $M_{\rm turn}$ is reduced exponentially. We use the default parameters in figure~\ref{fig:T_spin}.

The evolution of the spin temperature
can be summarized by dividing it into four phases.
First, the gas temperature and the CMB decoupled around $z \sim 200$. After the decoupling, the spin temperature well coupled with the gas temperature through the collisional coupling; thus $T_{\rm K} \simeq T_{\rm s} < T_{\rm CMB}$.
At the second phase, the spin temperature gradually approaches the CMB one
because the collisional interaction becomes weak due to the cosmic expansion. Thus three types of temperatures we considered obey the relation: $T_{\rm K} < T_{\rm s} \simeq T_{\rm CMB}$.
In the third phase, the first luminous objects start to emit the Ly-$\alpha$ photons and the Ly-$\alpha$ interaction becomes effective. Up to this stage, the gas temperature decreases adiabatically as the redshift decreases after the decoupling from the CMB one and the gas is cooler than the CMB. Thus, the spin temperature also gets lower than the CMB one, i.e., $T_{\rm K} \simeq T_{\rm s} < T_{\rm CMB}$.
As indicated in equation~(\ref{dTb}), the global 21-cm signal should be negative in this phase.
In other words, the 21-cm signals are observed as the absorption line.
After that, in the fourth phase, a large amount of X-ray photons can be emitted as a consequence of active star formations.
The X-ray photons heat the gas quickly in the IGM, and the gas temperature surpasses the CMB temperature.
Therefore, as the Ly-$\alpha$ coupling is still effective, the spin temperature also becomes higher than the CMB. At the time, the 21-cm signal is measured as an emission line. In the fiducial model, the ionization gradually starts by the UV photons emitted from high-$z$ galaxies after the beginning of X-ray gas heating. 
At this point, the global IGM 21-cm signal vanishes again.

\subsubsection{Status of the 21-cm global signal experiments \label{sec:status}}

\begin{figure}
\centering
\includegraphics[width=0.6\textwidth]{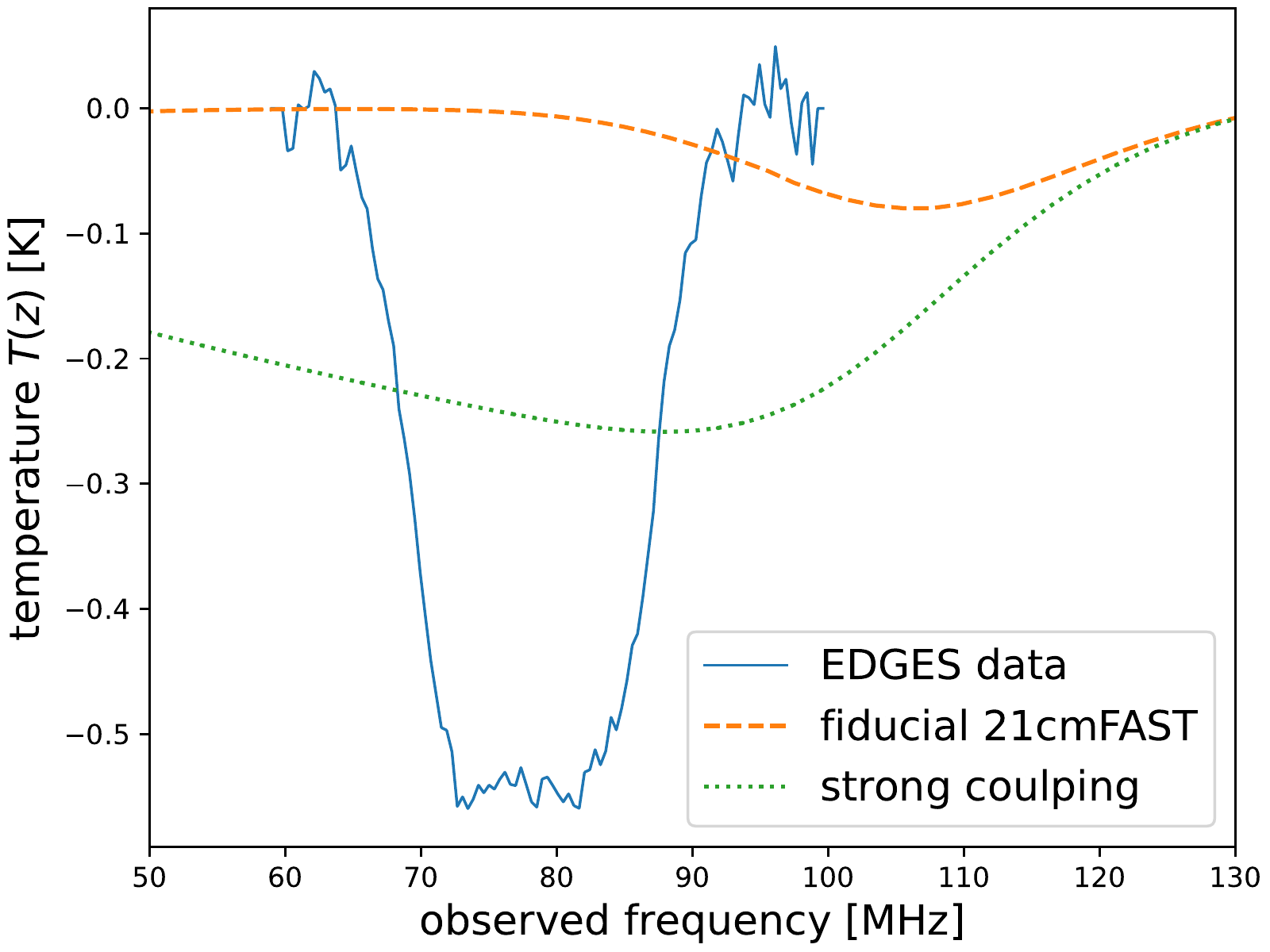}
\vspace{-0mm}
\caption{
Best-fit 21-cm absorption profiles, obtained in \cite{2018Natur.555...67B},
is plotted as the blue solid line. The orange dashed and green dotted lines represent the theoretical predictions of 21-cm global signal of the fiducial model of {\tt{21cmFAST}} \citep{2011MNRAS.411..955M,2020JOSS....5.2582M} and the model with strong coupling. 
}
\label{fig:edges}
\end{figure}

As mentioned in the introduction, many projects are working to measure the 21-cm global signal. Some have reported their results, and reionization and heating models have been tested against the observations of the 21-cm global signal such as EDGES high-band \citep{2017ApJ...847...64M}, LEDA \citep{2016MNRAS.461.2847B} and SARAS2 \citep{Singh2018SARASReionization2}.
Importantly, \cite{2018Natur.555...67B} has reported
the detection of the strong radio absorption signal around 78~MHz
with EDGES low-band as shown in figure~\ref{fig:edges}.
EDGES is a ground-based radio instrument
placed at Murchison Radio-astronomy Observatory,
and it has two different versions of the system;
a low-band instrument which is sensitive to the frequencies of 50--100~MHz and a high-band instrument for 100--200~MHz.
The blue solid line in figure~\ref{fig:edges} shows
the differential brightness temperature obtained by the EDGES low-band data
after subtracting the foreground emission with the four-term polynomial fitting \citep{2018Natur.555...67B}.

The reported brightness temperature in figure~\ref{fig:edges}
gives an absorption profile centered around 78~MHz,
and can be interpreted as the 21-cm signal redshifted at $z\sim17$,
which is to be produced by the Ly-$\alpha$ interaction and X-ray heating
mentioned above. 
However, the measured absorption profile has surprisingly deviated
from the theoretically expected ones
for the shape and depth of the absorption trough.
For comparison, we show a typical 21-cm global signal
and an extremely strong absorption signal in figure~\ref{fig:edges},
which are obtained by using \texttt{21cmFAST}.
Same as figure~\ref{fig:T_spin}, a prediction for the fiducial model, which is plotted by the orange dashed line, is calculated by adopting the default set of parameters in {\tt 21cmFAST}.
On the other hand, the strong absorption case with the green dotted line in figure~\ref{fig:edges} is computed by assuming $T_{\rm s} = T_{\rm K}$ at all redshifts, which corresponds to the dotted line in figure~\ref{fig:T_spin}.
As the gas temperature is determined by the adiabatic cooling, this extreme model demonstrates the strongest absorption with the observed frequency $<90$ MHz.
Even in such an extreme case, the expected absorption depth should be shallower than $\delta T_{\rm b} > -0.3$~K for 50--100 MHz and it hardly becomes deeper than $-0.3$~K in the standard scenario.
However, the observed absorption profile with EDGES is significantly deeper than $\delta T_{\rm b} = -0.5$~K.
We also note here that there have been some discussions on the interpretation/analysis regarding the EDGES results \citep{Hills:2018vyr,2018ApJ...858L..10D,
Bradley:2018eev,Singh:2019gsv,Spinelli:2019oqm}.
In any case, the anomaly suggested by EDGES measurements should be confirmed by the other instruments; SARAS3, SCI-HI, BIGHONS, LEDA, CTP, REACH, and PRI$^{\rm Z}$M. Recently, SARAS3 has reported their result and the best fitting absorption reported from EDGES has been rejected with $2\sigma$ confidence~\citep{2021arXiv211206778S}. Not only the single antenna measurement, but also some interesting methods have been proposed to measure the 21-cm global signal, such as the observation of the moon as a calibration source (\cite{2013AJ....145...23M,10.1093/mnras/stv746,2018MNRAS.481.5034M}) and using an array of closely-spaced antennas \citep{2020MNRAS.499...52M}. As the potential for systematic errors is unavoidable, it is important to use these other methods of 21-cm line observation.

Note that the reported absorption by the EDGES has prompted us to propose exotic models. These models predict non-standard signatures of the 21-cm global signal at the dark ages as well. Thus, observations at less than $50~\rm MHz$ are the interesting frontier to validate the models.
Such low-frequency observations allow us to conduct a purely cosmological analysis avoiding astrophysical uncertainties because no stars are expected to form during the dark ages.
However, the radio signal at less than $10~\rm MHz$ cannot be measured from Earth due to the reflection of Earth's ionosphere. Therefore, there are many projects to measure the 21-cm line at the dark ages from far-side of the Moon, such as Dark Ages Polarimeter PathfindER (DAPPER, \cite{2021arXiv210305085B}), Farside Array for Radio Science Investigations of the Dark ages and Exoplanets (FarSide, \cite{Burns:2021pkx}), Netherlands-China Low frequency Explorer (NCLE, \cite{2020AdSpR..65..856B}), and Lunar Crater Radio Telescope on the Far-Side of the Moon (LCRT\footnote{\url{https://www.nasa.gov/directorates/spacetech/niac/2020_Phase_I_Phase_II/lunar_crater_radio_telescope/}}, \cite{9438165}).
Ultimately, these observations would be expected to provide fruitful cosmological constraints from 21-cm line observations.

\subsubsection{Cosmology with 21-cm global signal}

As described above, the EDGES signal can be interpreted as the redshifted 21-cm signal
of~$15 \lesssim z \lesssim 20$. Although the depth of the absorption trough is significantly deeper than the one predicted in the standard cosmological scenario, the observed absorption signal might indicate that the spin temperature of neutral hydrogen gases was coupled well with the kinetic gas temperature, which was cooler than the CMB temperature around these redshifts. 
On the other hand, it may be also possible that scenarios with overcooling of the IGM gas or some extra radiation at radio frequency can explain the extremely strong absorption line.
Many works were performed to explain the anomalous signal by considering dark matter-baryon interaction \citep{Barkana:2018lgd,Munoz:2018pzp,Fialkov:2018xre,
Kang:2018qhi,Falkowski:2018qdj,Jia:2018csj,Sikivie:2018tml},
the nature of dark sector~\citep{Costa:2018aoy,Hill:2018lfx,Li:2018kzs},
production of photon at radio frequency to increase $T_{\rm CMB}$
\citep{Ewall-Wice:2018bzf,Fraser:2018acy,Yang:2018gjd,Pospelov:2018kdh,
Lawson:2018qkc,Moroi:2018vci,AristizabalSierra:2018emu,Fialkov2019SignatureSpectrum2,2020MNRAS.499.5993R}, and so on.

On the other hand, the EDGES absorption signal can be exploited to constrain cosmological heating sources since extra heating erases the absorption signal.
Furthermore, as the Ly-$\alpha$ radiation from high-$z$ objects affects the 21-cm global signal, the structure formation and the nature of density fluctuations can be constrained. Many authors have already employed the EDGES result to constrain several models.
Examples of the heating source include:
primordial magnetic fields \citep{2019MNRAS.488.2001M,Bera:2020jsg,Natwariya:2020ksr},
the Hawking evaporation of small Primordial Black Holes (PBHs)
\citep{Clark:2018ghm,Halder:2021rbq,Yang:2020egn},
the emission from the accretion disks around large PBHs~\citep{Hektor:2018qqw},
the decaying or annihilating dark matter
\citep{Clark:2018ghm,Cheung:2018vww,DAmico:2018sxd,
Liu:2018uzy,Mitridate:2018iag,Hektor:2018lec},
and so on.

The 21-cm global signal can be useful to constrain cosmological models that affect the structure formation (e.g., the timing of the switch-on of X-ray and Ly-$\alpha$ radiation from first objects depends on the shape of the matter power spectrum).
Examples of such models include warm dark matter models \citep{2018ApJ...859L..18S, 2018PhRvD..98b3011L, 2018PhRvD..98f3021S, 2018MNRAS.481L..69S}, the model of the primordial power spectrum at small scales~\citep{YoshiuraGSI,YoshiuraGSII,Munoz:2019hjh}.

As summarized above, various works have been performed to investigate
how the 21-cm global signal can probe various aspects of cosmology.
In sub-subsections~\ref{sec:Constraint from 21-cm global signal} and \ref{sec:21-cm global signal}, 
we discuss how the heating sources and models giving structure formation beyond the standard cosmology can be constrained by the EDGES result in some detail by taking two examples:
primordial magnetic fields~\citep{2019MNRAS.488.2001M} and the primordial curvature power spectrum~\citep{YoshiuraGSI,YoshiuraGSII}.

\subsection{21-cm fluctuation in IGM}

\subsubsection{Observables of 21-cm fluctuation \label{sec:21cm_fluc_IGM}}

In addition to the 21-cm line global signal, the fluctuation of the 21-cm line
has also useful information in the high-$z$ Universe. In the cosmological context, an important statistical quantity to characterize 21-cm fluctuation is the power spectrum.
While we have not detected the 21-cm power spectrum, radio interferometers have shown improvements of the upper limits on the power spectrum such as PAPER \citep{2019ApJ...883..133K}, GMRT \citep{Paciga2013AExperiment2}, MWA \citep{Barry2019ImprovingObservations2,2019ApJ...887..141L}, LOFAR \citep{2017ApJ...838...65P}, and HERA \citep{Deboer2017HydrogenHERA2} at EoR. Using the recent best upper limits from MWA 
\citep{Trott2020DeepObservations2}, LOFAR \citep{Mertens2020} and HERA Phase I \citep{HERAupper2021}, some astrophysical models have been disfavored and constrained (\cite{Greig2021MNRASMWA}, \cite{ Greig2021MNRASLOFAR} and \cite{HERAconst2021}).

Although the upper limits on the 21-cm power spectrum at the cosmic dawn were reported from MWA \citep{Ewall-Wice2016FirstHeating, Yoshiura2021a}, LWA \citep{Eastwood2019TheOVRO-LWA}, and LOFAR \citep{Gehlot2019TheLOFAR,2020MNRAS.499.4158G}, the current upper limits ($\ge 10^7 \rm mK^2$) at the cosmic dawn are not strong enough to constrain standard models. It should be worth mentioning that the EDGES low-band result suggested some exotic models which predict the enhancement of the power spectrum up to 3 orders of magnitudes (i.e. $10^6$ $\rm mK^2$, \cite{Fialkov2019SignatureSpectrum2}). Thus, further analysis could validate such extreme models. Note that significant improvements at the cosmic dawn are expected from up-coming instruments such as HERA \citep{Deboer2017HydrogenHERA2}, NenuFAR~\citep{8471648} and SKA1-LOW \citep{Koopmans2014TheArray}.

We here briefly introduce the formalism of the 21-cm power spectrum, based on \cite{Kohri:2013mxa}. We refer the readers to \cite{Kohri:2013mxa} for detail (and see also \cite{Mao:2008ug} for the analysis method). 
In \cite{Kohri:2013mxa}, they started the differential brightness temperature relative to the CMB.
In the expression of equation~(\ref{eq:IGMbrightness}), fluctuations of the brightness temperature $\delta_{\delta T_{b{\rm (IGM)}}}(\boldsymbol{x},z) =\delta  T_{b{\rm (IGM)} } (\boldsymbol{x},z) - \overline{\delta T}_{b{\rm (IGM)} } (z)$, with $\overline{\delta T}_{b{\rm (IGM)}}(z)$ being the spatially averaged 21-cm signal, is given by
\begin{eqnarray}
\label{eq:delta_Tb_IGM}
    \delta_{\delta T_{b{\rm (IGM)} }} (\boldsymbol{x},z)
    \approx \overline{\delta T}_{b{\rm (IGM)}}(z)\frac{1}{\bar{x}_{\rm HI}}
    \left( 1 - \bar{x}_i(z) \left(  1 +  \delta_x  ( \boldsymbol{x},z)  \right)\right) \nonumber \\
    \times\left( 1 + \delta (\boldsymbol{x},z) \right)
    \left( 1 - \delta_v (\boldsymbol{x},z)\right)
    - \overline{\delta T}_{b{\rm (IGM)}}(z)  \,,  
\end{eqnarray}
where $x_i$ is the ionization fraction, and $x_{\rm HI} = 1 - x_i$ is the neutral fraction.
$\delta_x (\boldsymbol{x},z)=  x_i (\boldsymbol{x},z) /\bar{x}_i(z)-1$ and $\delta (\boldsymbol{x},z)= n_{\rm H} (\boldsymbol{x},z)/ \bar{n}_{\rm H}(z) - 1 $ are the fluctuation of the ionization fraction and the number density of hydrogen, respectively. Barred quantities are the spatially averaged ones. Here, $\delta_v (\boldsymbol{x},z) = (1/aH) ({\rm d} v_r / {\rm d}r)$ is the term of peculiar velocity. 
Since we consider the signal from EoR, we have assumed the condition $T_{\rm s} \gg T_{\rm \rm CMB}$.
In the following discussions, we often omit the redshift $z$ in the arguments.
By using Fourier transformed fluctuation of 21-cm signal one can define the 21-cm power spectrum $P_{21} (\boldsymbol{k})$ in the $k$-space by 
\begin{equation}
    \left\langle \delta_{\delta T_{b{\rm (IGM)}}}^\ast (\boldsymbol{k}) \,\, \delta_{\delta T_{b{\rm (IGM)}}} (\boldsymbol{k}')  \right\rangle
    = (2\pi)^3 \delta^{3}_{\rm D}
    (\boldsymbol{k} - \boldsymbol{k}') P_{21} (\boldsymbol{k}) \,,
    \label{eq:21powerIGM}
\end{equation}
where $\langle\cdots\rangle$ denotes the ensemble average and $\delta^{3}_{\rm D}(\boldsymbol{k})$ is the three-dimensional Dirac delta function.
Since the peculiar velocity field can be related to the fluctuation of the hydrogen number density $\delta (\boldsymbol{k})$ in the $\boldsymbol{k}$-space as $\delta_v (\boldsymbol{k})= - \mu^2 \delta (\boldsymbol{k})$ with $\mu=\widehat{\boldsymbol{k}} \cdot \widehat{\boldsymbol{n}}$ representing the angle between the wave number $\boldsymbol{k}$ and the direction of the line-of-sight (LoS) $\boldsymbol{n}$, one can separate the contributions as 
\begin{equation}
     P_{21} (\boldsymbol{k})  
     =
     P_{\mu^0} (k)  + \mu^2 P_{\mu^2}(k) + \mu^4 P_{\mu^4}(k) \,,
\end{equation}
where \begin{eqnarray}
   P_{\mu^0} &=& \overline{\delta T}_{b{\rm (IGM)}}^2\left(  P_{\delta\delta} - 2  \bar{x}_{\rm i}/ \bar{x}_{\rm HI} P_{x\delta} + \bar{x}^2_{\rm i}/\bar{x}^2_{\rm HI} P_{xx} \right) \,, \nonumber \\
   P_{\mu^2} &=& 2\overline{\delta T}_{b{\rm (IGM)}}^2(  P_{\delta\delta} -  \bar{x}_{\rm i}/\bar{x}_{\rm HI} P_{x\delta} ) \,,
   \qquad
   P_{\mu^4} = \overline{\delta T}_{b{\rm (IGM)}}^2  P_{\delta\delta} \,.
\end{eqnarray}
Here, $P_{\delta\delta}$ and $P_{xx}$ are the power spectra for $\delta$ and $\delta_x$, respectively, and $P_{x\delta}$ represents their cross power spectrum. For the power spectrum involving the ionization fraction $P_{xx}$ and $P_{x\delta}$, the formulas fitted to some radiative transfer simulations are adopted in \cite{Kohri:2013mxa}.

The SKA might produce calibrated radio images given in angular and frequency spaces in which the position of the sky is represented by the vector 
\begin{equation}
   \boldsymbol{\Theta} = \boldsymbol{\Theta}_\perp + \Delta f \, \widehat{\boldsymbol{e}} \,,
\end{equation}
where $\Delta f$ is the difference of frequency from the centre frequency and  $\widehat{\boldsymbol{e}}$ is the unit vector along the LoS direction. Thus, we observe the power spectrum in the $\boldsymbol{u}$-space rather than the $\boldsymbol{k}$-space, where $\boldsymbol{u}$ is the Fourier dual of $\boldsymbol{\Theta}$. Using the comoving angular diameter distance, $d_{\rm A}(z)$, and $y(z) = \lambda_{21} (1+z)^2 /H(z)$ with $\lambda_{21}$ corresponds to the wavelength of 21-cm line, $\boldsymbol{\Theta}_\perp$ and $\Delta f$ can be related to the vector normal to the LoS direction $\boldsymbol{r}_\perp$ and the corresponding comoving distance $\Delta r_\parallel$ such as
\begin{equation}
    \boldsymbol{\Theta}_\perp = \frac{\boldsymbol{r}_\perp}{d_{\rm A} (z)} \,,
    \qquad
    \Delta f = \frac{\Delta r_\parallel}{y(z)} \,,
\end{equation}
The power spectra in $\boldsymbol{k}$- and $\boldsymbol{u}$-spaces can be related as 
\begin{equation}
    P_{21} (\boldsymbol{u}) = {d_{\rm A}^{-2}(z)  y^{-1}(z)} P_{21} (\boldsymbol{k}) \,.
    \label{eq:P_21 def}
\end{equation}

\subsubsection{21-cm signal from minihalos} \label{sec:21cm_minihalos}

Next, we give a brief review of the expected constraints
from the 21-cm signal from minihalos discussed in 
\cite{Sekiguchi:2017cdy}.
Minihalos are expected to exist in the dark ages/EoR as virialized objects with relatively low virial temperature $T_{\rm vir} < 10^4~{\rm K}$.
Because of such a low virial temperature, the collisional ionization is inefficient and then the inside of a minihalo would be filled with dense neutral gas.
Thus, in addition to the IGM fluctuations discussed in the previous subsection,
the inhomogeneous spatial distribution of minihalos
can be observed as the 21-cm neutral hydrogen fluctuations.

The virial temperature is given by~\citep{Barkana:2000fd}
\begin{equation}
T_{\rm vir} \sim 2 \times 10^4 \left( \frac{M}{10^8\, h^{-1} M_\odot} \right)^{2/3} \left( \frac{1+z}{10} \right)~{\rm K}.
\end{equation}
Thus, the typical mass of minihalos is roughly $\lesssim 10^8 ~ M_\odot$ and the abundance of minihalos depends on the matter density fluctuations at the corresponding scales, $k \gtrsim 100~{\rm Mpc}^{-1}$.
This means that observing minihalos through 21-cm line should be a powerful tool to probe the primordial fluctuations at smaller scales in comparison with other cosmological observations such as CMB which reveal the primordial fluctuations with $k \lesssim 1~{\rm Mpc}^{-1}$.

Like the 21-cm signal from IGM in equation~(\ref{eq:IGMbrightness}),
we can also describe the signal from minihalos by
the mean differential brightness temperature at redshift $z$ as~\citep{Iliev:2002gj}
\begin{equation}
\overline{\delta T}_{b{\rm (MH)}} (z) = \frac{(1+z)^4}{H(z)\nu_{21}}
\int^{M_{\rm max}}_{M_{\rm min}} {\rm d} M
\frac{{\rm d} n_{\rm MH}}{{\rm d} M} \Delta \nu_{\rm eff} A
\langle \delta T_{b{\rm (MH)}} \rangle~,
\end{equation}
where $\Delta \nu_{\rm eff}$ is the redshifted effective line-width given by $\Delta \nu_{\rm eff} = \left[ \phi(\nu_{21}) (1+z) \right]^{-1}$ with the intrinsic line profile $\phi(\nu_{21})$ for an optically thin minihalo, $A$ is the cross-section of a minihalo, and $\langle \delta T_{b{\rm (MH)}} \rangle$ is the brightness temperature of a minihalo averaged over the cross-section $A$.
The information of the small-scale primordial fluctuations could be imprinted in the above differential brightness temperature through the mass function of the minihalos ${\rm d}n_{\rm MH}/{\rm d}M$.

Furthermore, the distribution of minihalos is in general a biased tracer of the underlying matter density fluctuations, and thus the differential brightness temperature is biased as well.
By introducing the effective bias parameter $\beta (z)$,
we can write the 21-cm fluctuations from minihalos at a redshift $z$ in the LoS direction $\widehat{\boldsymbol{n}}$
as
\begin{equation}
\delta_{\delta T_{b{\rm (MH)}} } (\widehat{\boldsymbol{n}},z )
    = \overline{\delta T}_{b{\rm (MH)}}(z) \beta (z) \delta (\boldsymbol{x}=r(z) \widehat{\boldsymbol{n}},z),
    \label{eq:TBMH}
\end{equation}
where $r(z)$ is the comoving distance to the redshift $z$.
Thus, once the effective bias $\beta (z)$ is given appropriately,
similarly to the IGM case where the 21-cm power spectrum
is given by equation~(\ref{eq:21powerIGM}),
we can quantitatively evaluate the 21-cm power spectrum induced from the minihalos.

In the analysis performed in \cite{Sekiguchi:2017cdy}, we use
a truncated isothermal sphere
as the model of a minihalo and the Press-Schechter mass function, and set $M_{\rm max}$
and $M_{\rm min}$ in the integration to the virial mass
with the virial temperature $T_{\rm vir} = 10^4~{\rm K}$
and the Jeans mass, respectively.
We also have taken the redshift space distortions due to the peculiar velocity of
minihalos into account.
In the quantitative analysis, we use the angular power spectra of the 21-cm fluctuations:
\begin{equation}
    \langle a_{\ell m}^{\rm (MH)} (z) a_{\ell' m'}^{\rm (MH)} (z') \rangle
    = C_\ell^{\rm (MH)} (z,z') \delta_{\ell \ell'}\delta_{m m'},
\end{equation}
with
\begin{equation}
a_{\ell m}^{\rm (MH)} (z) = \int {\rm d} \widehat{\boldsymbol{n}} \delta_{\delta T_{b{\rm (MH)}}} (\widehat{\boldsymbol{n}}, z) Y_{\ell m}^\ast(\widehat{\boldsymbol{n}}).
\end{equation}

\subsection{21-cm signals from galaxies}

After the Universe was reionized, most neutral hydrogen (HI) would be found in dense systems inside galaxies, and consequently, we assume that the HI is a biased tracer of galaxy distributions. The observed 21-cm line emission from galaxies contains information on the redshift of galaxies, which can be used to reconstruct the three-dimensional distributions of underlying matters. This is called the HI galaxy redshift survey. This survey mode involves detecting the redshifted 21-cm emission from many individual galaxies above the flux threshold. The 21-cm line emission is quite weak, so we require a highly sensitive telescope such as SKA to perform large HI galaxy survey across a wide redshift range. Actually, the HI galaxy redshift survey delivered by SKA is expected to have the potential to be competitive with other next-generation cosmological experiments with different wavelengths. Basic prediction for the number density and corresponding bias of galaxies that will be detected by SKA HI galaxy surveys were made in \cite{Yahya:2014yva,Bull:2015lja}, using the following fitting functions
\begin{eqnarray}
	&&\frac{{\rm d}n_{\rm gal}(z)}{{\rm d}z}=10^{c_1}\,{\rm deg}^{-2}z^{c_2}{\rm exp} (-c_3z)
	\,,\\
	&&b_{\rm gal}(z)=c_4{\rm exp} (c_5z)
	\,.
\end{eqnarray}
Here the numerical values of the galaxy number density and bias prediction, $c_i$ ($i=1,\cdots ,5$), 
for the SKA Phase-1 and SKA Phase-2 specifications are presented in \cite{Bull:2015lja,Bacon:2018dui}.
The primary purpose of spectroscopic galaxy redshift surveys is generally to measure the three-dimensional
clustering of galaxies.
Because the galaxy clustering is measured as a function of redshift rather than comoving distance, we must
account for redshift space distortions caused by the peculiar velocity field of galaxies.
The power spectrum for the galaxy clustering at a given redshift and in the LoS direction $\widehat{\boldsymbol{n}}$ is given by
\begin{eqnarray}
	P_{\rm gal}(\boldsymbol{k},z)=\left( b_{\rm gal}(z)+f(z)\mu^2\right)^2 P_\delta (k,z)
	\,,
\end{eqnarray}
where $\mu =\widehat{\boldsymbol{k}}\cdot\widehat{\boldsymbol{n}}$, 
$f(z)$ denotes the linear growth rate of structure, and $P_\delta (k,z)$ is the isotropic matter power spectrum.
The HI galaxy survey delivered by the SKA will be able to provide accurate measurements
for the expansion history of the Universe and the growth rate of the large-scale structure, 
particularly the baryon acoustic oscillation and redshift space distortions.

In addition to the HI galaxy redshift survey, the intensity mapping of the 21-cm line has been proposed as
an innovative technique to probe the large-scale structure of the Universe. 
In order for the HI galaxy redshift survey to resolve individual galaxies, we need a sufficiently long integration 
time of the highly sensitive telescope.
On the other hand, the MID HI intensity mapping, instead of resolving individual galaxies, will 
measure the intensity of
the integrated 21-cm line emission of several unresolved galaxies in a reasonably large three-dimensional pixel.
Fluctuations in the observed intensity of this redshifted 21-cm line emission
are expected to follow fluctuations in the underlying matter density, which can be traced by the 21-cm line emitted galaxies.
The SKA will be capable of performing the MID HI intensity mapping surveys over $0<z<3$.
In terms of the brightness temperature, the average over the sky can be written as \citep{Bull:2014rha}
\begin{eqnarray}
	\overline{\delta T}_b(z)\approx\frac{3}{32\pi}\frac{hc^3A_{10}}{k_{\rm B}m_{\rm p}\nu_{21}^2}\frac{(1+z)^2}{H(z)}\Omega_{\rm HI}(z)\rho_{{\rm c},0}
	\,,
\end{eqnarray}
where the quantities $m_{\rm p}$, $\Omega_{\rm HI}$, and $\rho_{{\rm c},0}$ denote
the proton mass, the HI energy density fraction, and the critical density today, respectively.
Assuming that the signal is linear with respect to the underlying matter fluctuations, 
the total brightness temperature at a given redshift and in a direction $\widehat{\bf n}$ can be written as
\begin{eqnarray}
	\delta T_{b}(\widehat{\boldsymbol{n}},z)\approx\overline{\delta T}_b(z)\left( 1+b_{\rm HI}\delta +\frac{1}{aH}\frac{{\rm d}v_\parallel}{{\rm d}r}\right)
	\,.
\end{eqnarray}
The signal can be completely specified once we find a prescription for the HI density $\Omega_{\rm HI}$ and bias function $b_{\rm HI}$.
These can be obtained by making use of the halo mass function, halo bias, and the model
for the amount of HI mass in a dark matter halo~\citep{Santos:2015gra}.
The power spectrum for the brightness temperature delivered from the HI intensity mapping can be written as
\begin{eqnarray}
	P_{\rm IM}(\boldsymbol{k},z)=\overline{\delta T}_{b}^2(z)\left( b_{\rm HI}(k,z)+f(z)\mu^2\right)^2 P_\delta (k,z)
	\,.
\end{eqnarray}
The cross-correlation between the HI intensity mapping and the galaxy clustering depends on the galaxy bias $b_{\rm gal}$
and it can be used to mitigate systematic effects.

\subsection{Short summary}

As discussed above, several results have already been reported on the 21-cm global signal and fluctuations and in the future, further observational results are anticipated. In light of this, many works have been devoted to studying the potential abilities of the global signal and fluctuations of neutral hydrogen 21-cm line. 

So far cosmological information has been obtained mainly from observations of CMB and large-scale structure, which have successfully established the standard paradigm, the so-called $\Lambda$-CDM cosmology. Cosmological parameters such as energy densities of baryon, dark matter, and dark energy have also been accurately measured. However, there still remain several important issues to be unveiled, such as the actual mechanism of inflation, the nature of dark matter and dark energy, and so on.

Observations of CMB and large-scale structure have probed fluctuations on large scales which brought us significant knowledge on the evolution and present state of the Universe. However, we need to go beyond the current approach to sharpen our understanding of the Universe further, which can be possible using observations of 21-cm line, in particular, from the next generation telescope such as SKA. In the following sections, based on the 21-cm line signal presented above, we discuss how current and future observations of neutral hydrogen 21-cm line can probe many cosmological aspects to resolve various issues in modern cosmology. In the next section, we start our discussion with the issue of how the primordial Universe would be elucidated with observations of 21-cm line.

\section{Application to inflation \label{sec:inflation}}

\subsection{Current status}

The Universe is considered to have experienced a superluminal expansion at its very early stage, which is called inflation. Inflation can not only solve the problems in the standard big-bang cosmology such as the horizon and flatness problems, but also provide the origin of cosmic density fluctuations imprinted in CMB anisotropies, large-scale structure of the Universe, and so on. Indeed by using observations of CMB and large-scale structure, we can probe the inflationary Universe since the statistical nature of primordial density fluctuations depends on the inflation models.  

The properties of primordial fluctuations can be described by the correlation functions of the so-called curvature perturbation $\zeta$. One of the well-measured quantities is the primordial power spectrum of (scalar) density perturbation ${\cal P}_\zeta(k)$, which is defined by
\begin{equation}
    \left\langle \zeta (\boldsymbol{k}_1) \zeta (\boldsymbol{k}_2) \right\rangle
    =
    \frac{2\pi^2}{k_1^3} {\cal P}_\zeta (k_1)  (2\pi)^3 \delta^3_{\rm D} (\boldsymbol{k}_1 + \boldsymbol{k}_2 )  \,.
\end{equation}
Once we specify a model of inflation, one can calculate ${\cal P}_\zeta(k)$, which can be compared to the observed one. When one probes ${\cal P}_\zeta (k)$, it is commonly parametrized as 
\begin{eqnarray}
P_{\zeta} (k) = A_s (k_{\rm ref}) \left(\frac{k}{k_{\rm ref}}\right)^{n_{\rm s}-1 + \frac{1}{2}\alpha_s \ln (k/k_{\rm ref}) + \frac{1}{3!}\beta_s \ln^2 (k/k_{\rm ref})},
\label{eq:runnings}
\end{eqnarray}
where $ A_s (k_{\rm ref})$ is the amplitude
at the reference scale $k_{\rm ref}$, $n_{\rm s}$ is the spectral index, and
$\alpha_s$ and $\beta_s$ are the so-called running parameters. The amplitude $A_s (k_{\rm ref})$ and the spectral index $n_s$ are well measured by current CMB observations such as Planck and their values from the analysis of TT,TE,EE+lowE+lensing \citep{Akrami:2018odb} are 
\begin{eqnarray}
 \log (10^{10} A_s (k_{\rm ref}) ) &=& 3.044 \pm 0.014 \,, \\ 
 n_s &=& 0.9649 \pm 0.0042 \,,
\end{eqnarray}
where the quoted constraints are at 68~\% C.L. and the reference scale is taken as $k_{\rm ref}=0.05~ {\rm Mpc}^{-1}$. The running parameters are set to be zero to obtain the constraints.
When the running parameters are included in the analysis, Planck data gives the following limit:
\begin{eqnarray}
 \alpha_s &=& 0.002 \pm 0.010 \,, \\
 \beta_s &=& 0.010 \pm 0.013 \,.
\end{eqnarray}
Actually, CMB can only probe large-scale fluctuations, although the running parameters are constrained as mentioned above, these limits are not so severe enough to differentiate inflationary models. Therefore we need yet other observations to probe the scale-dependence of the primordial power spectrum more precisely. Indeed future observations of 21-cm line of neutral hydrogen can probe small-scale fluctuations, which would have 
a significant impact on constraining the running parameters. As we will review in the following, 21-cm fluctuations of neutral hydrogen from IGM/minihalos 
\citep{2013JCAP...10..065K,2017JCAP...05..032M,2018JCAP...02..053S}
and ultracompact minihalos \citep{2012PhRvD..85l5027B,2018JCAP...01..007E,2020MNRAS.494.4334F} can potentially measure the running parameter much more accurately than the current CMB limit\footnote{It has been also argued that other observations such as CMB spectral $\mu$ distortions \citep{2012PhRvD..86b3514D,2013JCAP...06..026K,2016PhRvD..93h3515C,2017JCAP...11..002K,2012ApJ...758...76C} and primordial black holes \citep{2009PhRvD..79j3511B,2009PhRvD..79j3520J,2019PhRvD.100f3521S}
can give further tight constraints on the running parameters.}. Even the 21-cm global signal obtained by EDGES can already reach the comparable limit with Planck, which will also be discussed in the following subsection. 

In addition to the primordial power spectrum, we can also probe higher-order statistics
such as primordial bi- and tri-spectra, which are respectively defined by
three- and four-point correlation functions of the curvature perturbation as 
\begin{eqnarray}
 && \left\langle \zeta (\boldsymbol{k}_1) \zeta (\boldsymbol{k}_2) \zeta (\boldsymbol{k}_3) \right\rangle \nonumber \\
&& \qquad \qquad 
= (2\pi)^3 B_\zeta (k_1, k_2, k_3) \delta_{\rm D}^3 (\boldsymbol{k}_1 + \boldsymbol{k}_2 + \boldsymbol{k}_3)  \,. \\ [8pt]
&& \left\langle \zeta (\boldsymbol{k}_1) \zeta (\boldsymbol{k}_2) \zeta (\boldsymbol{k}_3)\zeta (\boldsymbol{k}_4) \right\rangle \nonumber \\
    && \qquad \qquad 
= (2\pi)^3 T_\zeta (\boldsymbol{k}_1, \boldsymbol{k}_2, \boldsymbol{k}_3, \boldsymbol{k}_4) \delta_{\rm D}^3 (\boldsymbol{k}_1 + \boldsymbol{k}_2 + \boldsymbol{k}_3 + \boldsymbol{k}_4)  \,,
\end{eqnarray}
where $B_\zeta$ and $T_\zeta$ denotes the primordial bi- and tri-spectra, respectively. 
The functional forms of $B_\zeta$ and $T_\zeta$ depend on the generation mechanism of primordial fluctuations.
One of theoretically well-motivated forms is the so-called local-type non-Gaussianities, which are given by the following~\citep{Komatsu:2001rj,Okamoto:2002ik,Boubekeur:2005fj,Kogo:2006kh,Sasaki:2006kq}:
\begin{eqnarray}
&& B_\zeta (k_1,k_2,k_3) = \frac{6}{5} f_{\rm NL} \left( P_\zeta (k_1) P_\zeta (k_2) + 2~{\rm perms.} \right)~ \,,\label{eq:local bi} \\ [8pt]
&& T_\zeta (\boldsymbol{k}_1, \boldsymbol{k}_2, \boldsymbol{k}_3, \boldsymbol{k}_4) \nonumber \\
&& \qquad\qquad =  \tau_{\rm NL}  \left( P_\zeta (k_1) P_\zeta (k_2) P_\zeta (|\boldsymbol{k}_1+\boldsymbol{k}_3|) + 11~{\rm perms.} \right) \nonumber \\
&&\qquad\qquad + \frac{54}{25} g_{\rm NL}  \left( P_\zeta (k_1) P_\zeta (k_2) P_\zeta (k_3) + 3~{\rm perms.} \right)~,\label{eq:local tri}
\end{eqnarray}
where $f_{\rm NL}, \tau_{\rm NL}$ and $g_{\rm NL}$ are the non-linearity parameters which characterize the size of primordial bi- and tri-spectra for the loca-type non-Gaussianties. Planck data gives the constraints on these parameters as follows \citep{Planck:2013wtn,Akrami:2019izv}:
\begin{eqnarray}
f_{\rm NL} &=& -0.9 \pm 5.1 \qquad (68~\%~{\rm C.L.}) \,, \\ [8pt]
\tau_{\rm NL} &<& 2800 \qquad (95~\%~{\rm C.L.})  \,, \\ [8pt]
g_{\rm NL} &=& (-5.8 \pm 6.5) \times 10^4 \qquad (68~\%~{\rm C.L.}) \,.
\end{eqnarray}
Actually many inflationary models predict $f_{\rm NL} < {\cal O}(1)$, which indicates that current limit should be more improved to discriminate models of inflation.
Indeed, future observations of 21-cm fluctuations and galaxy surveys can probe $f_{\rm NL}$ down to ${\cal O}(0.1)$ level. Furthermore, one can also probe the non-linearity parameters for the trispectrum such as $\tau_{\rm NL}$ and $g_{\rm NL}$ much more precisely in future observations of 21-cm and galaxy surveys.

Besides the quantities mentioned above, testing adiabaticity (or isocurvature modes) of primordial fluctuations can also provide important implications for the inflationary scenario because any deviations from the adiabaticity indicate that multiple sources of fluctuations should exist.
Although current CMB observations such as Planck already give tight constraints on various isocurvature modes \citep{Akrami:2018odb}, observations of 21-cm line of neutral hydrogen that probe small-scale fluctuations can provide the information on qualitatively different aspects of isocurvature fluctuations which CMB cannot probe.
We also discuss this issue in the following subsection. 

Now in this section below, we discuss 21-cm global and fluctuation signals as a probe of the inflationary Universe, particularly focusing on the constraints on the running parameters, non-Gaussianities and the adiabaticity of primordial fluctuations in some detail.

\subsubsection{Constraint from 21-cm global signal}
\label{sec:Constraint from 21-cm global signal}

In \cite{YoshiuraGSI}, we investigated the impact of the primordial power spectrum
on the evolution of the 21-cm global signal and attempted to constrain the running parameters
based on the EDGES result rather than using the 21-cm power spectrum.
To investigate the effects of the primordial power spectrum to the 21-cm signal,
we use \texttt{21cmFAST}\footnote{We used an older version of {\tt 21cmFASTv1.2}. The older version is essentially the same that the later versions, although some parameters are removed and different assumptions (e.g. sub-grid recombination model) are adopted.} \citep{2011MNRAS.411..955M}. The variance of the density fluctuation is given as
\begin{eqnarray}
\sigma^2(R) &=& \int\frac{{\rm d}^3{\boldsymbol{k}}}{(2\pi)^3} P_{\rm m}(k) W^2(k,R),
\label{eq:sigma}
\end{eqnarray}
using the matter power spectrum at $z=0$, $P_{\rm m}(k)$, and the scale in real space, $R$.
As the window function $W(k, R)$, the real space top-had filter is applied in the \texttt{21cmFAST}. Through the value of $\sigma$, the primordial power spectrum affects the halo mass function.
For example, a large positive value of the runnings enhances the matter power spectrum at small scales and promotes structure formation earlier, and vice versa. As a result, the evolution of 21-cm global signal shifts to higher or lower redshift. We note that the {\tt 21mFAST} has various key parameters such as the minimum virial temperature, $T_{\rm vir}$, which controls the minimum mass of halo emitting energetic photons, and the number of X-ray photons per solar mass in stars, $\zeta_{\rm X}$. We first fixed these astrophysical parameters\footnote{We ignored parameters related to ionization in our analysis, but the effect of such parameters should be small since we discuss the 21-cm signal well before the reionization.}. 

\begin{figure}
\centering
\includegraphics[width=0.6\textwidth]{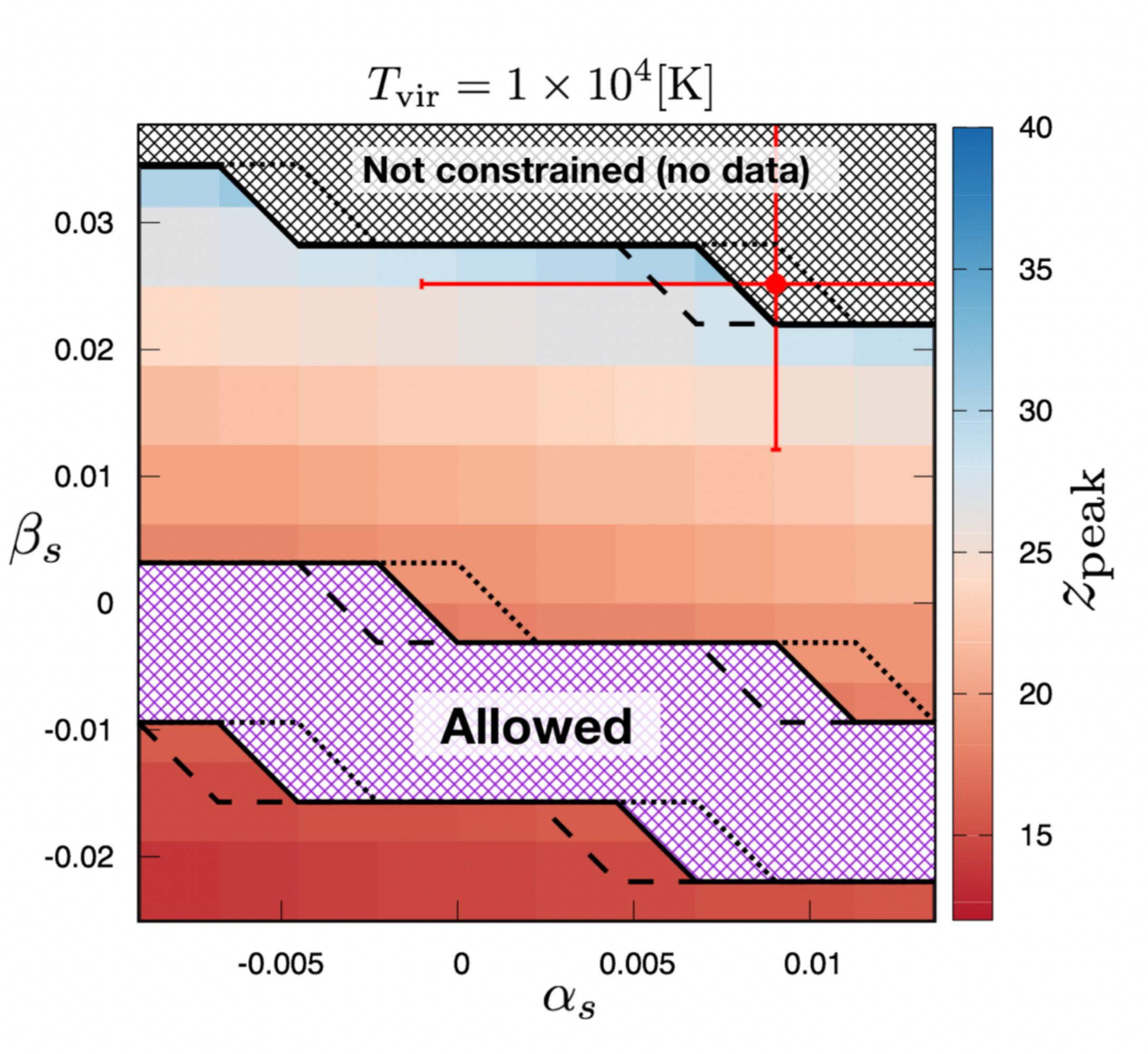}
\vspace{-0mm}
\caption{
Constraints from the EDGES result for the cases with $T_{\rm vir}=10^4 {\rm K}$.
The color indicates the peak redshift of absorption trough, $z_{\rm peak}$.
The dotted, solid and dashed lines are the result for $n_{\rm s}$ = 0.9530, 0.9586 and 0.9642.
These values correspond to central and 1 $\sigma$ limits from Planck \citep{2016A&A...594A..20P}.
Red point is best fit value of Planck and the error bars are 1 $\sigma$ error.
This figure is reproduced from \cite{YoshiuraGSI}}
\label{fig:construnning}
\end{figure}

As mentioned in the previous section, the EDGES signal shows a strong absorption line
which would be impossible to explain without exotic models and/or some unknown systematics.
However, given the fact that the EDGES has reported
no signal with root-mean-square of $25$~mK at $z\leq 14$ and $z \geq 22$,
we judge a model as disfavored if the absorption is stronger than $100$~mK
in the redshift ranges of $z \leq 14$ and $z \geq 22$.
If the peak of absorption trough is at $z>27$,
we do not constrain such models because the EDGES has not observed these redshifts. 
Figure~\ref{fig:construnning} shows the constraints
on the running parameters for a fixed astrophysical model.
The hatched region indicated as ``Allowed" is allowed one.
The disfavored region below the allowed one in figure~\ref{fig:construnning}
has a lower amplitude of the primordial power spectrum at small scales,
and the absorption trough is settled at $z \le 14$.
On the other hand, the region above the allowed one gives
too early structure formation and Ly-$\alpha$ coupling becomes effective earlier. As a result, the absorption trough appears at $22 \le z \le 27$.
Interestingly, the models consistent with Planck~\citep{2016A&A...594A..20P}\footnote{The latest result of Planck \citep{2020A&A...641A..10P}
is statistically consistent with the allowed models.}
can be disfavored from the EDGES signal.
This indicates the 21-cm line is a powerful measurement to probe the primordial power spectrum. 
In the original paper, we investigated the impact of astrophysical parameters on the running constraints. The running parameters degenerate with $T_{\rm vir}$ and $\zeta_{\rm X}$
and the constraints in figure~\ref{fig:construnning}
depend on the assumption for astrophysical parameters.
The degeneracy between running parameters and astrophysical parameters
will be removed by combining the future observations of the 21-cm power spectrum
and galaxy luminosity function.

In \cite{YoshiuraGSI}, we model the primordial power spectrum using running parameters.
Although this is a commonly assumed form, it is not necessarily the case when a non-smooth primordial power spectrum is realized.  
For example, the power spectrum may be affected by local feature of inflation potential,
double inflation(-like) models can give an enhanced power spectrum only on small scales 
\citep[e.g.][]{2017JCAP...09..020K}.
Furthermore, multi-fields models also can generate primordial fluctuations
in which large and small scale ones are generated from different fields \citep[e.g.][]{2016JCAP...04..057E}.
In the light of these considerations,
in \cite{YoshiuraGSII}, we model the primordial power spectrum such that it is enhanced by a factor of $p$ instead of describing it with the extrapolation from the large-scale amplitude such as the one given in equation~(\ref{eq:runnings}). Thus, the primordial power spectrum is given as
\begin{eqnarray}
P_{\zeta}(k) = A_{s}(k_{\rm ref}) \left(\frac{k}{k_{\rm ref}}\right)^{n_s-1} p(k).
\end{eqnarray}
A parameter $p(k)$ is a factor over the scales of $10<k<1000 ~{\rm Mpc}^{-1}$ and unity at other scales.
As the same manner as \cite{YoshiuraGSI}, given the form of the primordial power spectrum, 
by comparing the predicted 21-cm global signal with the EDGES non-detection result \citep{2017ApJ...847...64M,2018Natur.555...67B},
we found the tight bounds on the enhancement factor as $2<p<8$ although astrophysical parameters are fixed  in the analysis. 
It should be emphasized that the 21-cm global signal has a potential to give lower bound on the $p(k)$,
which has not been obtained in previous works.
Thus, the 21-cm line observation will be a unique probe of the primordial power spectrum.
It would also be worth mentioning that we here only use the EDGES non-detection results,
however, once future global signal observations
reveal precise properties of the global spectrum,
the constraints should become tighter than those obtained by these works.

\subsubsection{Constraint from $21$-cm fluctuations}

In the previous section, we discussed constraints on the primordial power spectrum from the 21-cm global signal. We can also constrain it by using observations of 21-cm fluctuations. Since the 21-cm signal can be generated from IGM and/or minihalos, below we briefly discuss expected constraints on the primordial power spectrum using 21-cm fluctuations from IGM and minihalo separately. 

By using the formalism described in sub-subsection~\ref{sec:21cm_fluc_IGM}, one can discuss how precisely observations of 21-cm fluctuations from IGM can constrain the primordial power spectrum, particularly its detailed scale-dependence via the running parameters $\alpha_s$ and $\beta_s$. In \cite{Kohri:2013mxa}, the Fisher matrix analysis for the power spectrum $P_{21} (\boldsymbol{u})$, which is defined in equation~(\ref{eq:P_21 def}), is adopted and the redshift range of $z=6.8 - 10$ are considered. Since this redshift range would include the reionization era, we also need the information on power spectra for ionization fraction $P_{xx}$ and $P_{x\delta}$, which are taken into account by assuming some specific form motivated to be matched with some radiative transfer simulations. Details of the analysis can be found in \cite{Kohri:2013mxa}.   It has been shown that the running parameters can be probed with the precision of $\delta \alpha_s = {\cal O}(10^{-3})$ and $\delta \beta_s = {\cal O}(10^{-3})$ by combining futuristic observations of CMB (such as COrE-like observations \citep{Finelli:2016cyd,Delabrouille:2017rct}) and SKA. In table~\ref{tab:running_IGM}, 1$\sigma$ uncertainties expected from some combinations of future observations are shown.

\begin{table}[ht]
    \centering
    \begin{tabular}{l|ccc}
    \hline 
         & $\delta n_s$ & $\delta \alpha_s$ & $\delta \beta_s$ \\ \hline \hline 
        COrE-like & $2.13 \times 10^{-3}$ & $2.43 \times 10^{-3}$ & $4.47 \times 10^{-3}$ \\ 
   COrE-like+SKA & $1.34 \times 10^{-3}$ & $1.85 \times 10^{-3}$ & $1.22 \times 10^{-3}$ \\ 
   COrE-like+Omniscope & $5.54 \times 10^{-4}$ & $1.00 \times 10^{-3}$ & $6.87 \times 10^{-4}$ \\ \hline
    \end{tabular}
    \vspace{2mm}
    \caption{1$\sigma$ uncertainties for $n_s, \alpha_s$ and $\beta_s$ expected from future observations of CMB and 21-cm fluctuations \citep{Kohri:2013mxa}.}
    \label{tab:running_IGM}
\end{table}

\begin{figure}[htbp]
  \begin{center}
   \hspace{0mm}\scalebox{2.0}{\includegraphics{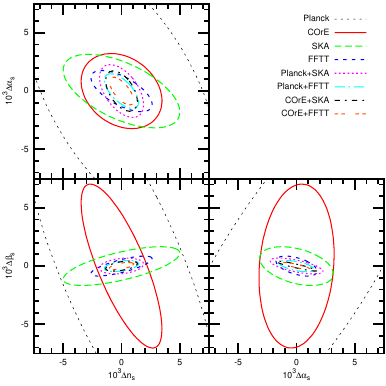}}
  \end{center}
  \vspace{5mm}
  \caption{Expected constraints on the running parameters of the spectral index of the primordial power spectrum from observations of 21-cm signals from minihalos in combination with the CMB, taken from \cite{Sekiguchi:2017cdy}.
  }
  \label{fig:zmin06}
\end{figure}

Expected constraints on the primordial power spectrum from 21-cm fluctuation generated from minihalos have also been investigated in \cite{Sekiguchi:2017cdy} where the angular power spectra for 21-cm fluctuations discussed in sub-subsection~\ref{sec:21cm_minihalos} are adopted to obtain the constraints. For the details of the analysis with minihalos, we refer the readers to \cite{Sekiguchi:2017cdy} (see also \citealt{Iliev:2002gj,Sekiguchi:2013lma,Sekiguchi:2014wfa,2014PhRvD..90h3003S}).
 In figure~\ref{fig:zmin06} which is taken from \cite{Sekiguchi:2017cdy},
we show expected constraints on the spectral index and its runnings
from the future observations of 21-cm signals from minihalos
in combination with CMB for several observations. Here we take the minimum redshift $z_{\rm min}$ until when minihalos can be observed to be $z_{\rm min} = 6$.  We also show 1$\sigma$ uncertainties expected from the combinations of future observations for the case of $z_{\rm min}= 6$ in table~\ref{tab:running_minihalos}. For cases of other minimum redshift, see~\cite{Sekiguchi:2017cdy}. Although the signals in 21-cm fluctuations can have contributions both from IGM and minihalo, in the works mentioned in this section, the case of either IGM or minihalo signal being dominant is discussed.  Although which signal dominates the signal depends on the reionization model, redshift range and so on, the signal from IGM or minihalos would give a similar sensitivity for the spectral index $n_s$ and the running parameters $\alpha_s$ and $\beta_s$, which is more precise than the current observational bounds obtained from Planck.
\begin{table}[htbp]
    \begin{center}
    \begin{tabular}{l|ccc}
    \hline 
         & $\delta n_s$ & $\delta \alpha_s$ & $\delta \beta_s$ \\ \hline \hline
        COrE-like+SKA & $1.2 \times 10^{-3}$ & $1.6 \times 10^{-3}$ & $0.39 \times 10^{-3}$ \\ 
   COrE-like+FFTT (Omniscope) & $0.95 \times 10^{-3}$ & $1.1 \times 10^{-3}$ & $0.28 \times 10^{-3}$ \\ \hline
    \end{tabular}
    \vspace{2mm}
    \end{center}
    \caption{1$\sigma$ uncertainties for $n_s, \alpha_s$ and $\beta_s$ expected from future observations of CMB and 21-cm fluctuations by making use of the minihalos~\citep{Sekiguchi:2017cdy}.}
    \label{tab:running_minihalos}
\end{table}

\subsection{Isocurvature perturbations \label{sec:iso}}

As we have discussed in the previous section, the observations of the 21-cm fluctuations
from dark ages/EoR could be powerful to probe the primordial curvature perturbations
at smaller scales compared to other cosmological observations such as CMB and galaxy surveys.
We can also test the adiabaticity of the primordial density fluctuations at such smaller scales, i.e., isocurvature modes with 21-cm fluctuations~\citep{Sekiguchi:2013lma,Takeuchi:2013hza,Gong:2017sie,Kadota:2020ybe}.

Although CMB observations such as Planck severely constrain the amplitudes of isocurvature fluctuations on large scales \citep{Akrami:2018odb}, they cannot well constrain them on small scales. Power spectra of isocurvature fluctuations are usually parametrized in the same manner as the adiabatic ones as 
\begin{equation}
    {\cal P}_{\rm iso}^{(i)} (k) = A_{\rm iso}^{(i)} (k_{\rm ref})
    \left( \frac{k}{k_{\rm ref}} \right)^{n_{\rm iso}^{(i)}-1}\,,
\end{equation}
where $A_{\rm iso}^{(i)} (k_{\rm ref})$ and $n_{\rm iso}^{(i)}$ are the amplitude and spectral index for the isocurvature mode~$i$, respectively. When the power spectrum is (almost) scale-invariant, such isocurvature fluctuations are severely constrained on large scales from CMB as mentioned above. However, when the power spectrum is blue-tilted (i.e., $n_{\rm iso}^{(i)} > 1 $), the amplitude can get large on small scale while keeping the large-scale amplitude insignificant and such isocurvature fluctuations are not constrained by CMB. Models of blue-tilted isocurvature modes are discussed in the context of baryogenesis \citep{Yokoyama:1990xv}, primordial black holes (PBHs) \citep{Yokoyama:1995ex,Gong:2017sie}, axion \citep{Fairbairn:2017sil,Dai:2019lud,Kadota:2020ybe}, and so on. When such blue-tilted isocurvature fluctuations exist, they dominate the matter power spectrum on small scales, and then the structure formation at high redshifts are enhanced, which affects the 21-cm fluctuation signals. 

In \cite{Sekiguchi:2013lma,Takeuchi:2013hza}, it is argued that blue-tilted matter (CDM/baryon) isocurvature fluctuations can be probed/constrained by future observations of 21-cm fluctuations from minihalos such as SKA and more futuristic ones like FFTT (Omniscope), in which the basic formalism to calculate 21-cm fluctuations from minihalos are the same as the one discussed in sub-subsection~\ref{sec:21cm_minihalos} although the root-mean-square fluctuations of differential brightness temperature are adopted in the analysis of \cite{Sekiguchi:2013lma,Takeuchi:2013hza}. A similar analysis focusing on PBHs has been done in \cite{Gong:2017sie}. In \cite{Kadota:2020ybe}, expected constraints on axion isocurvature fluctuations in a scenario where  Peccei-Quinn symmetry is broken after inflation have been studied by adopting the angular power spectrum as discussed in sub-subsection~\ref{sec:21cm_minihalos}. In these studies, it has been demonstrated that 21-cm fluctuations from minihalos can be a powerful tool to probe isocurvature fluctuations which are enhanced on small scales. 

21-cm fluctuations can also be used to differentiate CDM and baryon isocurvature fluctuations \citep{Kawasaki:2011ze} (see also \cite{Barkana:2005xu}, and in the context of the so-called compensated isocurvature mode, see \cite{Gordon:2009wx}). It is well known that CDM and baryon isocurvature fluctuations give the same shape for the CMB power spectrum and hence CMB observations can only put constraints on the sum of CDM and baryon isocurvature fluctuations. 
However, the evolution of density fluctuations is in principle different for CDM and baryon isocurvature initial conditions. Since observations of 21-cm fluctuation basically probe fluctuations of baryon, which enable us to differentiate CMB and baryon isocurvature modes. In \cite{Kawasaki:2011ze}, it has been investigated to what extent the differentiation is possible assuming a futuristic observation of 21-cm fluctuation such as FFTT which can probe the signal at high redshifts as $z \ge 30$. It was shown that, although it is in practice difficult to differentiate these modes when isocurvature fluctuations have (almost) a scale-invariant power spectrum, we can differentiate CDM and baryon isocurvature modes when the power spectra are very blue-tilted \citep{Kawasaki:2011ze}, which shows that 21-cm fluctuations can be a very unique probe of primordial density fluctuations.

\subsection{Primordial non-Gaussianities}
 

In the standard inflation paradigm, the primordial curvature perturbations
have been originated from the quantum fluctuations of light scalar fields including
the inflaton and/or spectator fields.
Based on this picture, the generated primordial perturbations have to be Gaussian at the leading order.
However, due to the interaction of each field and the fact that even Einstein gravity has non-linearities, at the next-leading order they would acquire non-Gaussian features and thus
testing such a non-Gaussian nature of the primordial perturbations observationally
can give us the hints for what the inflationary model/the gravity theory in the high energy scales are.

There have been lots of works about constraining primordial non-Gaussianities
by making use of the future observations of 21-cm fluctuations~\citep{Cooray:2006km,2007ApJ...662....1P,Joudaki:2011sv,Tashiro:2011br,Tashiro:2012wr,Chongchitnan:2012we, Chongchitnan:2013oxa, DAloisio:2013mgn, Munoz:2015eqa}. 
Here, we focus only on the so-called local-type non-Gaussianities
(see equations~(\ref{eq:local bi}) and (\ref{eq:local tri})) which can be useful to probe the super-horizon evolution of the primordial curvature perturbations,
and this type of primordial non-Gaussianity has been extensively discussed in the
context of the cosmic structure formation, because it can give a characteristic and significant impact
in the halo/galaxy distribution on large scales, as we will see later.



\subsubsection{Use of IGM during dark ages}

In \cite{Cooray:2006km,2007ApJ...662....1P,Munoz:2015eqa}, to probe the primordial non-Gaussianities from the 21-cm observations, the authors focused on the bispectrum of the 21-cm fluctuations during the dark ages ($z \sim 30$ -- $200$), when any ionizing sources are expected to be absent and hence the neutral fraction of IGM is considered to be unity. As given in equation~(\ref{eq:delta_Tb_IGM}), the 21-cm fluctuations are roughly proportional to the matter density fluctuations at the leading order. Through the Poisson equation, the matter density fluctuations are linearly coupled to the primordial curvature perturbations. As we have shown, the non-linearity parameter $f_{\rm NL}$ represents the amplitude of the bispectrum of primordial curvature perturbations. Thus, the bispectrum of $\delta_{\delta T_{b({\rm IGM})}}$ should be proportional to the non-linearity parameter $f_{\rm NL}$.

They have estimated the angular bispectrum of 21-cm fluctuations, also including the secondary non-linear effect from the gravitational evolution, and performed the Fisher analysis for a cosmic-variance-limited experiment observing the full sky. They conclude that the bispectrum of the 21-cm fluctuations during the dark ages has a potential to give a tight constraint on the non-linearity parameter $f_{\rm NL}$ as $\delta f_{\rm NL} = 1.3~(0.23)$ for the experiment with arcminute (one tenth of arcminute) resolution and assuming a single narrow redshift slice at $z=50$. By making use of the tomographic method with integrating all the information from $z=30$ to $100$, the $1\sigma$ uncertainties will be much improved as $\delta f_{\rm NL} = 0.12~(0.03)$ for the experiment with a bandwidth of $\Delta \nu = 1~(0.1)~{\rm MHz}$~\citep{Munoz:2015eqa}.
Note that the current constraint on $f_{\rm NL}$ obtained from Planck is $-6 < f_{\rm NL} < 4.2$ (68\%~CL)~\citep{Akrami:2019izv}.
The standard single-field slow-roll inflation model predicts a small non-Gaussianity of $f_{\rm NL} < O(0.01)$ and hence the observational limit at this level will be an ultimate test for the standard inflation model. The observation of the 21-cm signals during the dark ages, $z \sim 30$ -- $200$, is actually difficult using the ground based telescopes due to the reflection of signals by the Earth's ionosphere below 40 MHz.
The future projects to measure the 21-cm line at the dark ages from the far-side of the Moon such as DAPPER~\citep{2021arXiv210305085B}, FarSide~\citep{Burns:2021pkx,Burns:2021ndk}, NCLE~\citep{2020AdSpR..65..856B}, and LCRT~\citep{9438165} could do it when it is realized. Such moon based instruments can avoid not only the ionospheric effects but also radio frequency interference which is one of the severe systematics of the 21-cm observation.

\subsubsection{Use of IGM during EoR}

As the other way to constrain the primordial non-Gaussianities by 21-cm observations, \cite{Joudaki:2011sv} discussed the effect of non-Gaussianity on the fluctuations of the ionized fraction $\delta_x$. There have been lots of discussions about hunting for the primordial non-Gaussianities from the large-scale clustering of halos/galaxies, the so-called scale-dependent bias (see, e.g. \cite{Dalal:2007cu}).
Basically, this effect is caused by the non-linear coupling between the short and long wavelength modes. In the case where the initial density fluctuations are purely Gaussian, the large-scale clustering of the biased objects such as halos/galaxies can be described by a scale-independent bias parameter, $b_{\rm G}$, as $\delta_{\rm biased} \simeq b_{\rm G} \delta_{\rm m}$ with $\delta_{\rm m}$ being the underlying density fluctuations.
On the other hand, in the case with the local-type primordial non-Gaussianity, it could induce the large-scale modulation of the small-scale variance of density fluctuations due to the non-linear coupling, and then the scale-dependence, even on large scales, would arise in the bias parameter. For the simplest case, that is, the constant $f_{\rm NL}$, the induced scale-dependent part of the bias is given by $\propto f_{\rm NL}/k^2$. Thus, the precise measurement of the large-scale clustering of the biased objects enables us to obtain the constraint on $f_{\rm NL}$.

In the 21-cm observations, roughly speaking, the clustering of ionized regions is considered to be corresponding to that of the biased objects and thus the power spectrum of $\delta_x$ can be expressed as $P_{xx} \simeq b_x^2 P_{\delta \delta}$.
Based on the above fact, the bias parameter $b_x$ can be simply parameterized as $b_x = b_0 + b_1 f_{\rm NL}/k^2$ with $b_0$ and $b_1$ being constant parameters in the existence of the local-type $f_{\rm NL}$. In \cite{Joudaki:2011sv}, the authors performed the Fisher matrix analysis and concluded that SKA can constrain $f_{\rm NL}$ with an accuracy of order $10$.
The constraint can be improved by FFTT (Omniscope) as $\delta f_{\rm NL} \sim 0.6$.
\cite{Mao:2013yaa} has also discussed the potential of 21-cm fluctuations from EoR and shown that SKA can achieve $\delta f_{\rm NL} \sim 3$ by making use of tomographic multi-frequency observations, which is comparable to the current constraint obtained from the Planck CMB observation.

\subsubsection{Use of minihalos}

\begin{figure*}[htbp]
\centering 
   \includegraphics[width=.9\textwidth]{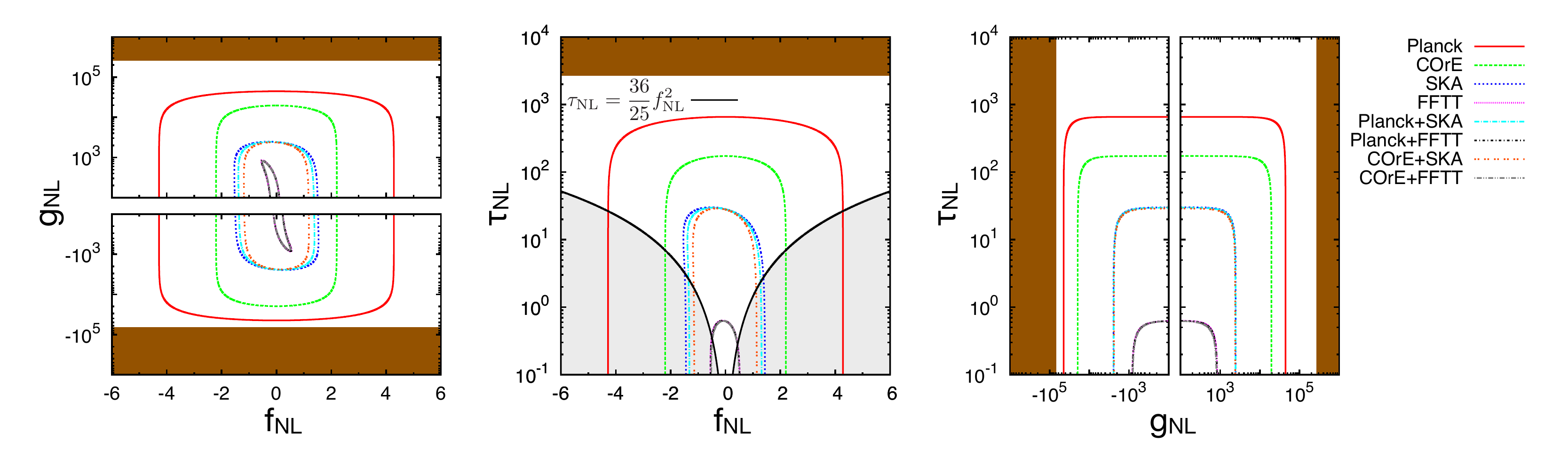}
\caption{\label{fig:fnlgnl} Expected 2-dim constraints on the non-linearity parameters assuming $z_{\rm min}=6$, taken from \cite{Sekiguchi:2018kqe}.
}
\end{figure*}

\begin{figure}[htbp]
\centering 
   \includegraphics[width=.5\textwidth]{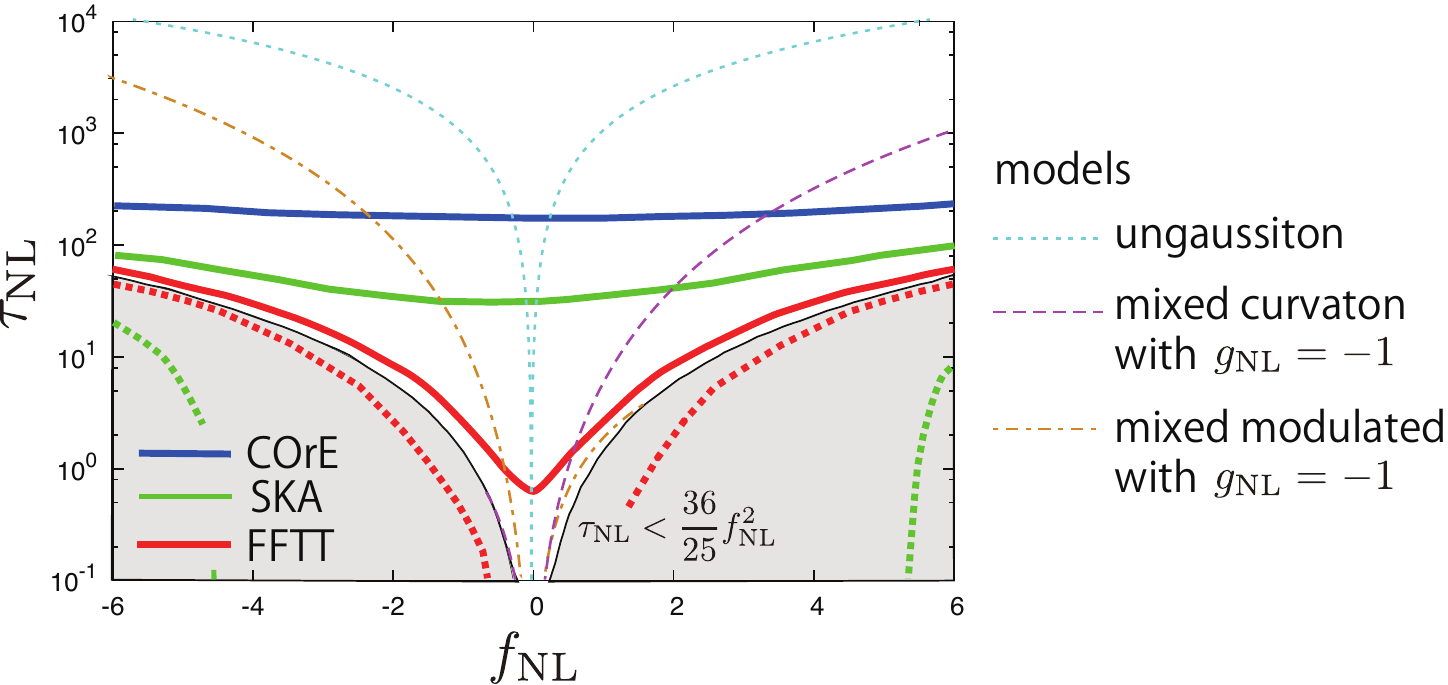}
\caption{\label{fig:consistency} 
Regions where the consistency relation $\tau_{\rm NL} = (36/25) f_{\rm NL}^2$ can be excluded at 1$\sigma$, taken from \cite{Sekiguchi:2018kqe}.
}
\end{figure}

Here let us focus on \cite{Sekiguchi:2018kqe} where we have discussed the potential of minihalos for constraining the primordial non-Gaussianities. Basically, minihalos are also considered to be biased objects and hence we can describe the clustering of minihalos by introducing the bias parameter $b_{\rm MH}$, which can be included in the effective bias parameter $\beta (z)$ introduced in equation~(\ref{eq:TBMH}) as
\begin{equation}
    \beta (z) 
    = \frac{\int dM \frac{dn}{dM}{\mathcal F}(z,M) b_{\rm MH}(k,M,z)}
    {\int dM \frac{dn}{dM}{\mathcal F}(z,M)}~,
\end{equation}
where ${\mathcal F}(z,M)$ is a flux from a single minihalo with the mass $M$ at the redshift $z$.
Then the power spectrum of the spatial fluctuations in the minihalo number density is roughly given by
\begin{equation}
    P_{\rm MH} (k, z, z') \propto \beta (k,z) \beta (k,z') P_{\delta\delta}~.
\end{equation}
In the analysis done in \cite{Sekiguchi:2018kqe}, in addition to the contribution from $f_{\rm NL}$, we have also included the effect of the higher-order local-type non-Gaussianity, in particular, primordial trispectrum which can be characterized by two parameters $g_{\rm NL}$ and $\tau_{\rm NL}$.
In such a case, formally, we have~\citep{Sekiguchi:2018kqe} (see, e.g. \cite{Gong:2011gx,Yokoyama:2012az}, for the usual halo power spectrum)
\begin{equation}
    P_{\rm MH} (k, z, z') \propto \left[\beta_0(z)\beta_0(z') + \frac{\Delta \beta (z)}{k^2} \beta_0(z') + \frac{\Delta \beta (z')}{k^2} \beta_0(z)+\frac{36}{25}\tau_{\rm NL} \frac{\beta_f(z)\beta_f(z')}{k^4} \right] P_{\delta\delta}~,
    \label{eq:minihalo power spectrum}
\end{equation}
with
\begin{equation}
    \Delta \beta (z) = \beta_f(z) f_{\rm NL} + \beta_g(z) g_{\rm NL}~.
\end{equation}
As can be seen in this equation, for a single redshift observation, that is, the case with $z'=z$, the contributions from the non-linearity parameters $f_{\rm NL}$ and $g_{\rm NL}$ are completely degenerate. Thus, in our Fisher matrix analysis, we assume the observations with the multiple redshifts and obtain the expected $1\sigma$ errors for the non-linearity parameters as in table~\ref{tab:nonG_minihalos} and in figure~\ref{fig:fnlgnl}.

\begin{table}[htbp]
    \centering
    \begin{tabular}{l|ccc}
    \hline 
         & $\delta f_{\rm NL}$ & $\delta g_{\rm NL}/10^3$ & $\delta \tau_{\rm NL}$ \\ \hline \hline
        COrE-like+SKA & $1.1$ & $2.2$ & $27$ \\ 
   COrE-like+FFTT~(Omniscope) & $0.48 $ & $0.75$ & $0.58$ \\ \hline
    \end{tabular}
    \caption{1$\sigma$ uncertainties for $f_{\rm NL}, g_{\rm NL}$ and $\tau_{\rm NL}$ expected from future observations of CMB and 21-cm fluctuations by making use of the minihalos~\citep{Sekiguchi:2018kqe}.}
    \label{tab:nonG_minihalos}
\end{table}

Once we achieve the constraints on non-linearity parameters at this level,
it should be powerful enough to probe inflationary models
of generating local-type primordial non-Gaussianities such as the curvaton model, modulated reheating scenario and so on~(see, e.g. \cite{Suyama:2010uj} for various models) as can be seen in figure~\ref{fig:consistency}.

\subsubsection{Use of galaxies}

Galaxy survey can access the distribution of matter in three dimensions, thus having access to a large number of modes than accessible to CMB experiments. In the linear regime, one of the most distinctive effects of the local-type primordial non-Gaussianities in the HI galaxy survey and MID HI intensity mapping survey is the scale-dependent enhancement of the large-scale clustering of galaxies due to the non-linear coupling caused by the primordial non-Gaussianities. The large-scale asymptotes of the bias should be modified as \citep{Dalal:2007cu,Matarrese:2008nc}
\begin{equation}
	b_{\rm gal/HI}(z)\to b_{\rm gal/HI}(z)+f_{\rm NL}\Delta b_f(z,k)
	\,,
\end{equation}
where the non-Gaussian correction to the Gaussian bias factor is given by
\begin{equation}
	\Delta b_f(z,k)=\frac{3H_0^2\Omega_{{\rm m},0}D_+(z_*)(1+z_*)b_f(z)}{k^2T(k)D_+(z)}
	\,,\label{eq:bias correction}
\end{equation}
where $D_+(z)$ and $T(k)$ denote the linear growth rate, matter transfer function, respectively.
$z_*$ represents an arbitrary redshift at the matter dominated era.
Here $b_f(z)$ is the function depending on the relation between the clustering feature of galaxy density field and linear density field.
For instance, when we adopt the halo bias prescription and employ the Press-Schechter mass function for the universal mass function, 
we have $b_f(z)=\delta_{\rm c}(b_{\rm gal/HI}(z)-1)$ with $\delta_{\rm c}$ being the critical linear density for spherical collapse.
Since the large-scale behavior of the transfer function ${\cal M}\propto k^2$, we analytically estimate the scaling relation of
the bias on large scales: $\Delta b_f\propto k^{-2}D_+(z)^{-1}$, which implies that the bias correction is prominent on the largest
cosmological scales and at higher redshift, which are accessible by a large-area galaxy clustering survey 
delivered by the SKA, using either the 21-cm line emission/absorption~\citep{Camera:2013kpa} or the radio continuum emission 
\citep{Raccanelli:2014kga,Raccanelli:2016fmc,Yamauchi:2014ioa} of galaxies.
With the SKA Phase-2 HI galaxy redshift survey, it should be possible to reach $f_{\rm NL}$ close to unity~\citep{Camera:2014bwa}.
This is an important threshold to distinguish between single-field and many multi-field inflationary scenarios.
Measuring the scale-dependent bias on very large scales requires extremely large cosmological volumes 
to reduce error bars.
However, even the next-generation galaxy surveys will not be able to bring
$\delta f_{\rm NL}$ below unity using the single tracer.
In order to beat cosmic variance one needs to take advantage of the multi-tracer technique~\citep{Seljak:2008xr} (see also \citealt{Yamauchi:2016wuc} for biased bispectrum)
to break the important threshold $\delta f_{\rm NL}\sim 1$ \citep{Yamauchi:2014ioa,Ferramacho:2014pua}.

Moreover, different shapes of higher-order primordial spectra can be linked to different mechanism for 
generating non-Gaussian features of the primordial fluctuations.
Considering the local-type primordial non-Gaussianity, the primordial trispectrum can be characterized 
by additional two parameters $g_{\rm NL}$ and $\tau_{\rm NL}$.
Although the simplest single source model predicts $\tau_{\rm NL}=((6/5)f_{\rm NL})^2$, even in a generic situation 
there is an universal relation $\tau_{\rm NL}\geq ((6/5)f_{\rm NL})^2$ \citep{Suyama:2007bg}. 
We should note that the non-Gaussian correction formula equation~(\ref{eq:bias correction}) is not valid when
the nontrivial $g_{\rm NL}$ and $\tau_{\rm NL}$ are considered.
In this case, the general expression including the higher-order non-linear parameters is given by
\begin{equation}
	P_{\rm gal}(k,z)
		=\biggl[
			b_{\rm gal}^2(z)+2b_{\rm gal}(z)\Bigl\{ f_{\rm NL}\Delta b_f(k,z)+g_{\rm NL}\Delta b_g(k,z)\Bigr\}
			+\frac{36}{25}\tau_{\rm NL}\Delta b_f^2(k,z)
		\biggr] P_\delta (k,z)
	\,,
\end{equation}
where $\Delta b_g$ denotes the non-Gaussian correction due to $g_{\rm NL}$.
Although its explicit expression is not shown here, $\Delta b_g$ is in proportion to
the skewness of the smoothed density field.
When the single source model with negligibly small $g_{\rm NL}$ is considered, 
the above expression can reduce to the well-known formula equation~(\ref{eq:bias correction}).
A detection of the higher-order non-linearity parameters and the confirmation of the consistency inequality
should be the target in future experiments.  
The constraining power of the SKA Phase-2 on the higher-order non-linearity parameters is several tens of times severer
than that from the current CMB observations~\citep{Yamauchi:2015mja,Yamauchi:2021nsf}, hence the SKA can detect
the consistency inequality in the wide parameter region. 
Moreover, an important observation is that the combination of the SKA with future optical observations such as Euclid~\footnote{http://www.euclid-ec.org}
\citep{2013LRR....16....6A} offers a unique opportunity to exclude the inflationary paradigm itself if we find the breaking of the consistency inequality.
\\

In addition to the amplitude of the primordial bispectrum, its shape encodes 
much physical information about the primordial Universe.
Different shapes can be linked to different mechanisms for generating primordial non-Gaussianities 
by precise large-scale structure measurements.
However, the scale-dependent clustering due to non-local-type primordial non-Gaussianities
in the power spectrum for biased objects is shown to be irrelevant because the scale-dependence due to the non-local-type one in the bias is given as $\Delta b\propto k^0, k^{-1}$ for the equilateral-
and orthogonal-type, respectively, and is too weak to detect \citep{Matsubara:2012nc,Schmidt:2010gw}.
Therefore, to constrain the non-local-type primordial non-Gaussianities, we should consider
the higher-order correlation functions for the galaxy clustering, such as the galaxy bispectrum.
The bispectrum for the biased objects should be generated from the late-time non-linear gravitational
evolution of the density fluctuation even in the case that the primordial fluctuation obeys the Gaussian statistics
and such the contributions strongly depend on the underlying cosmological dynamics~\citep{Yamauchi:2017ibz,Yamauchi:2021nxw}. 
It is shown that the galaxy bispectrum for the galaxy clustering 
due to the non-local-type and the higher-order primordial non-Gaussianities 
is enhanced on large scales~\citep{Sefusatti:2007ih}.
Actually, \cite{Yamauchi:2016wuc} shows that 
the measurement of galaxy bispectrum 
by the SKA HI galaxy surveys 
can reach the constraints
on the equilateral- and orthogonal-type primordial non-Gaussianities 
to the level severer than current 
one which has been obtained by CMB observations.

\section{Application to PMFs}
\label{sec:PMFs}
\subsection{Observations of cosmological magnetic fields}

Various observations have revealed the existence of magnetic fields in the cosmological structure including galaxies, clusters of galaxies and large-scale filamentary structure~\citep{2013A&ARv..21...62D, 2017ARA&A..55..111H, 2018ApJS..234...11H}.
Interestingly, in the last decade, it has been recognized that observations of high-energy TeV photons emitted by distant blazars can be explained by the existence of magnetic fields in the void regions, as one of the plausible scenarios~\citep{2010Sci...328...73N,2011MNRAS.414.3566T,2012ApJ...747L..14V,2011APh....35..135E,2012ApJ...744L...7T,2013ApJ...771L..42T,2014MNRAS.445L..41T,2015PhRvL.115u1103C}.
The suggested lower bound on the magnetic fields is $10^{-20}$--$10^{-15}\; {\rm Gauss}$ with their coherent length, $\lambda_{B} \gtrsim 0.1\; {\rm Mpc}$.
Note that there are some debates on other interpretations of TeV blazars measurements from magnetic fields \citep{2012ApJ...752...22B,2013ApJ...770...54M,2012ApJ...758..102S,2013arXiv1311.6752S,2019MNRAS.489.3836A}.
Revealing the origin of such cosmic magnetic fields is one of the big challenges in modern cosmology.

An interesting scenario for the origin of cosmic magnetic fields is assuming that seed fields, the so-called primordial magnetic fields (PMFs), are generated in the early universe, especially before the epoch of cosmic recombination (see e.g., \citealt{2013A&ARv..21...62D,2016RPPh...79g6901S,2001PhR...348..163G} for reviews).
Seed fields associated with galaxies and clusters of galaxies are amplified through the dynamo mechanism to the currently observed amplitudes. Besides, the intergalactic magnetic fields implied by the blazars observations can be interpreted as remnants of seed fields, and thus, seed fields have a potential to explain the observed magnetic fields.
One of the representative scenarios for generating PMFs is inflationary magnetogenesis.
According to these scenarios, PMFs are generated from quantum fluctuations as well as curvature perturbations and primordial gravitational waves.
However, as long as the electromagnetic action is Maxwell's theory which is invariant under the conformal transformation, electromagnetic fields generated via vacuum fluctuations are too tiny as the origin of magnetic fields in the cosmological structure.
In order for PMFs to be generated during inflation, one requires theories beyond the standard model of particle physics breaking a conformal invariance, for instance, a new interaction between electromagnetic fields and any other fields (see e.g., \citealt{1988PhRvD..37.2743T,1992ApJ...391L...1R} for pioneering works).
On the basis of this scenario, several recent works have presented the successful models for inflationary magnetogenesis~\citep{2016EL....11519001D,2014JCAP...10..056C,2019JCAP...09..008F}.
Concerning the other possibilities of the origin of PMFs, the models in the post-inflation era have been proposed, e.g., based on the Harrison mechanism~\citep{1970MNRAS.147..279H,2005PhRvL..95l1301T,2011MNRAS.414.2354F,2015PhRvD..91l3510S,2016PhRvD..93j3536F}, or the bubble collision/turbulence in the cosmological phase transitions~\citep{1991PhLB..265..258V,1997PhRvD..55.4582S,2012ApJ...759...54T,2019JCAP...09..019E}.
Therefore, understanding the PMFs not only gives keys to the origins of magnetic fields in the cosmological structures but also provides fruitful information about the physics in the early universe.

As of now, no conclusive evidence of PMFs has been found yet, but a lot of works have tackled to investigate the nature of PMFs, by constraining the PMF strength from cosmological observations, for instance, Big Bang Nucleosynthesis light element abundance, CMB and galaxy surveys.
More interestingly, it has been known that PMFs affect the thermal evolution and dynamics of the IGM gas during the cosmic dawn and EoR, and thus, future redshifted 21-cm observations, e.g. SKA, are expected to be a good probe of PMFs.
In this section, we first introduce the basic properties of PMFs, such as the statistical properties and time evolution in subsection~\ref{sec:PMFs_basics}.
Next, we review the effects of PMFs on the thermal evolution and dynamics of the IGM gas during the dark ages in subsection~\ref{sec:PMFs_IGM}.
Finally, we summarize the relevant studies on the constraint on PMFs from future 21-cm observations in subsection~\ref{sec:PMFs_obs}.

\subsection{Basics of PMFs}
\label{sec:PMFs_basics}
Magnetic fields suffer the cosmological expansion.
When magnetic fields evolve adiabatically, the amplitude of magnetic fields, $B(t, \boldsymbol{x})$, decays proportional to $a^{-2}(t)$ with a scale factor $a(t)$ at a cosmic time $t$.
In analysing the PMFs evolution in the expanding Universe, it is useful to introduce comoving magnetic fields, $B(\boldsymbol{x})$, which is scaled to the strength at the present epoch, $B(\boldsymbol{x}) = B(t, \boldsymbol{x})/a^2(t)$.
In the following discussions, the strength of magnetic fields $B$ represents the comoving value unless explicitly stated otherwise.

Generally, PMFs are assumed to be generated in stochastic processes, and are often assumed to be Gaussian random fields.
In this case, the statistical nature of PMFs can be fully captured by the power spectrum as
\begin{equation}
\langle B_{i}(\boldsymbol{k}) B^{*}_{j}(\boldsymbol{k'})\rangle = (2\pi)^{3}\delta^{3}_{\rm D}(\boldsymbol{k}-\boldsymbol{k'}) \frac{\delta_{ij} - \widehat{k}_{i}\widehat{k}_{j}}{2}P_{B}(k) ~,
\label{eq:B-power-def}
\end{equation}
where $\widehat{k}_i$ is the $i$-th component of the normalized wave number vector $\widehat{\boldsymbol{k}}$, $\delta^{3}_{\rm D}$ is the Dirac delta function, and $B_i (\boldsymbol{k})$ is the Fourier component of PMF, which is defined by $\boldsymbol{B}(\boldsymbol{k}) = \int {\rm d}^3 x ~{\rm e}^{i\boldsymbol{k}\cdot \boldsymbol{x}} \boldsymbol{B}(\boldsymbol{x})$~.
Note that in equation~(\ref{eq:B-power-def}), we assume that PMFs are nonhelical fields.

In the context of the cosmological analysis, it is often used the following power-law type power spectrum of PMFs, which are suggested by various theoretical models of magnetogenesis.
Therefore, we adopt a power-law type power spectrum as
\begin{equation}
P_B(k) = A_B k^{n_B} \quad ({\rm for}\hspace{3mm} 
k \leq k_{\rm cut}), 
\label{eq:PowerB}
\end{equation}
where the parameters, $A_B$ and $n_B$, represent the amplitude and the scale dependence
of the PMF power spectrum, respectively.
We further introduce the cutoff scale of PMFs $k_{\rm cut}$ in equation~(\ref{eq:PowerB}).
This is because the PMF energy dissipates on small scales due to the radiative diffusion effect before the recombination epoch.

It is known that the CMB photons are diffused by the baryon-photon interaction before recombination, and as a result, the CMB temperature anisotropies are suppressed on small scales at $k \gtrsim 0.3~{\rm Mpc}^{-1}$. This mechanism is called the diffusion damping. Similarly, the energy density of PMFs on small scales is also diffused by the baryon-photon interaction. This diffusion scale is given by the random walk length determined with the Alfv\'en velocity, $v_{\rm A} \equiv c B_0 / \sqrt{4 \pi \rho_{{\rm b},0} a(t)}$ where $\rho_{{\rm b}}$ is the baryon density and the subscript $0$ denotes the present value.
Accordingly, the comoving cutoff scale induced by the direct cascade process is given by~\citep{1998PhRvD..57.3264J,1998PhRvD..58h3502S}
\begin{eqnarray}
\left(\frac{2\pi~{\rm Mpc}^{-1}}{k_{\rm cut}}\right)^2
&=& \frac{v_{\rm A}^2}{\sigma_{\rm T}}
\int^{t_{\rm rec}}_0 \frac{{\rm d}t}{a^2 n_{\rm e}} \nonumber \\
&\simeq& \left[1.32 \times 10^{-3} 
\left(\frac{B_n}{1~{\rm nG}}\right)^2
\left(\frac{\Omega_{\rm b} h^2}{0.02}\right)^{-1}
\left(\frac{\Omega_{\rm m} h^2}{0.15}\right)^{1/2}
\right]^{2/(n_B+5)}~,
\label{cutoff}
\end{eqnarray}
with the recombination time $t_{\rm rec}$.
Note that this comoving cutoff scale is time independent in the matter
dominated epoch.

Since we are interested in the strength of PMFs in real space,
instead of using $A_B$, it is helpful to introduce the strength smoothed on the scale $\lambda$,
\begin{equation}
B_\lambda^2 = \int \frac{{\rm d}^3 k}{(2\pi)^3}
~{\rm e}^{-k^2 \lambda^2} P_B (k) 
= \frac{A_B}{4\pi^2 \lambda^{n_B+3}}~\Gamma \left(\frac{n_B+3}{2}\right).
\label{eq:smoothedB}
\end{equation}
Here we use a Gaussian function for the smoothing window function.

Nowadays, many authors have studied the effects of PMFs on the cosmological observations such as CMB fluctuations~\citep{2004PhRvD..70d3011L,2004PhRvD..70l3507G,2008PhRvD..78b3510F,2009MNRAS.396..523P,2011MNRAS.411.1284T,2010PhRvD..81d3517S,2013PhRvD..88h3515B,2016A&A...594A..19P,2019MNRAS.490.4419S,2021JCAP...03..093M}
and the large-scale structures of the universe~\citep{2012SSRv..166....1R,2012JCAP...11..055F,2012PhRvD..86d3510S,2012MNRAS.424..927T,2014JCAP...03..027C}.
However, using CMB fluctuations and/or large-scale structure of the universe, one cannot expect stronger constraints on the amplitude of PMFs than of the order of a few nano-Gauss, because its energy density corresponds to the level of the CMB fluctuations (see e.g., \citealt{2001PhRvD..65b3517C}):
\begin{equation}
\frac{B^{2}/(8\pi)}{\rho_{\rm CMB,0}} \simeq 0.95\times 10^{-5} \left( \frac{T_{\rm CMB,0}}{2.725\, {\rm K}}\right)^{-4} \left(\frac{B}{10\, {\rm nG}} \right)^{2} ~.
\label{eq:mag-cmb}
\end{equation}
This relation implies that magnetic fields with several tens nano-Gauss contribute to the perturbations in the CMB anisotropy on the order of $10^{-5}$ which is the same level contribution from the curvature perturbations~(Recently, \citet{2019PhRvL.123b1301J} analyzes the magnetohydrodynamic (MHD) effect on CMB temperature anisotropy in detail, and gives a stringent upper limit on the amplitude of PMFs as $0.047 \; {\rm nG}$).
In other words, it is difficult to get the stronger constraint on PMFs through similar ways as to probe the curvature perturbations in the CMB observations.

On the other hand, measuring the thermal evolution of IGM gas would offer an interesting opportunity to explore PMFs with the energy density smaller than equation~(\ref{eq:mag-cmb}) because the thermal energy of the IGM is much smaller than the CMB energy density, and it makes us possible to observe the impacts of PMFs on IGM gas.
Ongoing and upcoming 21-cm observations have the potential to probe the thermal history of the IGM in high redshifts. More interestingly, the PMFs do not only affect thermal history but also generate additional density fluctuations on smaller scales than 1~Mpc.
The measurement of the density fluctuations on such smaller scales is one of the new frontiers in cosmology.
Since future high-redshift 21-cm observations are sensitive to both IGM density and temperature, it can be expected to lead us to obtain the tighter constraint on PMFs.

\subsection{Impact of PMFs on the baryon physics in the dark ages}
\label{sec:PMFs_IGM}

Even after the universe became neutral at the recombination epoch, baryon gas can tightly couple with PMFs through the residual ionized particles.
Therefore, the MHD approximation is valid in the dark ages.
PMFs coupled with baryon fluids mainly provide two MHD effects on the IGM: one is the heating on the IGM and the other is the generation of the density fluctuations~\citep{1978ApJ...224..337W, 2005MNRAS.356..778S}.
In this subsection, we discuss the impact of PMFs on the IGM thermal history and density fluctuations after the recombination epoch.

\subsubsection{Impact on the IGM thermal history}

Because of the coupling between PMFs and baryons through the MHD effect, the PMF energy dissipates and heats the IGM.
There are two dissipation processes after the recombination epoch: the ambipolar diffusion and the decaying turbulence.

The ambipolar diffusion is the energy dissipation process due to the collision between neutral and ionized particles~\citep{1992pavi.book.....S}.
After the recombination, IGM consists mostly of neutral baryon particles and residual ionized particles.
In such a partially ionized medium with magnetic fields, ionized particles are forced on by the Lorentz force while neutral particles are not.
This difference of motions induces the relative velocity between ionized and neutral particles. However, since these particles frequently collide in a cosmological time scale, the induced relative velocity is damped and heat up the gas temperature by the collision.
As a consequence, the ambipolar diffusion transfers the magnetic field energy to the thermal energy of baryons through the energy exchange of the relative velocity between neutral and ionized particles.

\cite{1998PhRvD..58h3502S} has pointed out that, before the recombination epoch, PMFs cannot induce turbulence on small scales because of the strong radiative viscosity.
After recombination, baryon-photon interactions rapidly decrease, and thus the radiative viscosity becomes less effective.
Resultantly, the magnetic Reynolds number is getting higher, and the turbulence motion can be produced in the MHD fluid.
Under such situations, the non-linear interactions between turbulences for different scales cause the energy cascade from larger vortex motions into small ones.
The kinetic energy due to such small-scale vortex motions and the magnetic energy coupled to baryons are dissipated into the thermal energy of the MHD fluid.
This mechanism is called the decaying MHD turbulence, and the time evolution of magnetic energy in the flat-space has been investigated well by numerical simulations~\citep{1996PhRvD..54.1291B, 2001PhRvE..64e6405C, 2004PhRvD..70l3003B}.
\citet{2005MNRAS.356..778S} has applied these results into the discussion of the effects on the IGM after the recombination.

The heating rates of the ambipolar diffusion, $\dot{Q}_{\rm AD}$, and the decaying turbulence processes, $\dot{Q}_{\rm DT}$, are given by~\citep{2005MNRAS.356..778S}
\begin{eqnarray}
\dot{Q}_{\rm AD} &=& \frac{|(\nabla \times \boldsymbol{B}) \times \boldsymbol{B}|^2}{16\pi^2 \xi \rho_{\rm b, 0}^2 a^4(t)} \frac{1-x_{\rm e}}{x_{\rm e}}~,
\label{gamma_pmf_ad} \\
\dot{Q}_{\rm DT} &=& \frac{3w_B}{2} H \frac{|\boldsymbol{B}|^2}{8\pi a^4(t)}
\frac{\left[\ln(1+t_{\rm d}/t_{\rm rec})\right]^{w_B}}{\left[\ln(1+t_{\rm d}/t_{\rm rec}) + \ln(t/t_{\rm rec})\right]^{1+w_B}}~,
\label{gamma_pmf}
\end{eqnarray}
where we define $\xi = 1.9 (T_{\rm K}/1 K)^{0.375} \times 10^{14}~{\rm cm}^3~{\rm g}^{-1}~{\rm s}^{-1}$ by the drag coefficient, $t_{\rm d} = (k_{\rm cut} v_{\rm A})^{-1}$ by the time-scale of decaying turbulence, and $w_B = 2(n_B+3)/(n_B+5)$ by the time-dependence of decaying turbulence.
Here, the operator $\nabla$ stands for the derivative with respect to the comoving coordinates.

The background thermal evolution of the IGM with these heating processes of the PMFs are calculated from
\begin{eqnarray}
\frac{{\rm d}T_{\rm K}}{{\rm d}t} = -2H T_{\rm K}
+ \frac{x_{\rm e}}{1+x_{\rm e}} \frac{8\rho_{\rm CMB} \sigma_{\rm T}}{3m_{\rm e} c} (T_{\rm CMB} - T_{\rm K})
+ \frac{2}{3k_{\rm B} n_{\rm b}} (\langle \dot{Q}_{\rm AD} \rangle + \langle \dot{Q}_{\rm DT}\rangle) ~,
\label{T_gas}
\end{eqnarray}
where the heating rates, $\langle \dot{Q}_{\rm AD} \rangle$ and $\langle \dot{Q}_{\rm DT}\rangle$, include the ensemble averages, $\langle |(\nabla \times \boldsymbol{B}) \times \boldsymbol{B}|^2 \rangle$ and $\langle |\boldsymbol{B}|^2 \rangle$, respectively.
In computing these ensemble averages, we exploit the expression of the PMF power spectrum given in equation~(\ref{eq:PowerB}).

The PMF heating in equation~(\ref{T_gas})
leads to the gas temperature rise enough to increase the residual ionization fraction $x_e$.
The evolution of the ionization fraction can be evaluated in
\begin{equation}
\frac{dx_{\rm e}}{dt} =
D \left[-\alpha_{\rm e} n_{\rm b} x_{\rm e}^2 + \beta_{\rm e} (1-x_{\rm e})
\exp \left(-\frac{3 E_{\rm ion}}{4 k_{\rm B} T_{\rm CMB}}\right)\right]
+ \gamma_{\rm e} n_{\rm b} x_{\rm e} (1-x_{\rm e})~.
\label{x_e} 
\end{equation}
Here, the first and second terms in the square brackets of the right-hand side represent the collisional-recombination and photo-ionization processes having coefficients $\alpha_e$ and $\beta_e$, depending on the kinetic gas temperature, respectively~\citep{2000ApJS..128..407S}.
The factor $D$ is the suppression factor due to the Ly-$\alpha$ resonance, which is determined both by the redshift rate of Ly-$\alpha$ photon and by the two-photon decay rate~(for details, see e.g.~\citealt{2000ApJS..128..407S}).

The last term on the right-hand side represents the effect of collisional ionization with thermal electrons.
In the standard recombination history, this term has a negligible contribution. 
However, when PMF heating exists, the gas temperature can increase up to $\sim 10^4~{\rm K}$, depending on the PMF strength. In this case, the collisional ionization is the major process for ionization.
The collisional ionization coefficient, $\gamma_e$, is a function of the kinetic gas temperature $T_{\rm K}$~\citep{1997ADNDT..65....1V}.
We refer the readers to \citet{2015MNRAS.451.2244C} for detailed discussions.

Recently, \citet{2009ApJ...692..236S, 2019MNRAS.488.2001M} found that neglecting the PMF dissipation of the magnetic field evolution leads to the overestimation of the 21-cm signals.
Thus, incorporating this contribution into the analysis, we write the evolution of the PMFs with the dissipation processes by
\begin{equation}
\frac{{\rm d}\rho_{\rm mag}}{{\rm d}t}
= -4H\rho_{\rm mag}
-\langle \dot{Q}_{\rm AD} \rangle - \langle \dot{Q}_{\rm DT}\rangle~.
\label{E_mag}
\end{equation}

\begin{figure}[h!]
\centering
\includegraphics[width=0.6\textwidth]{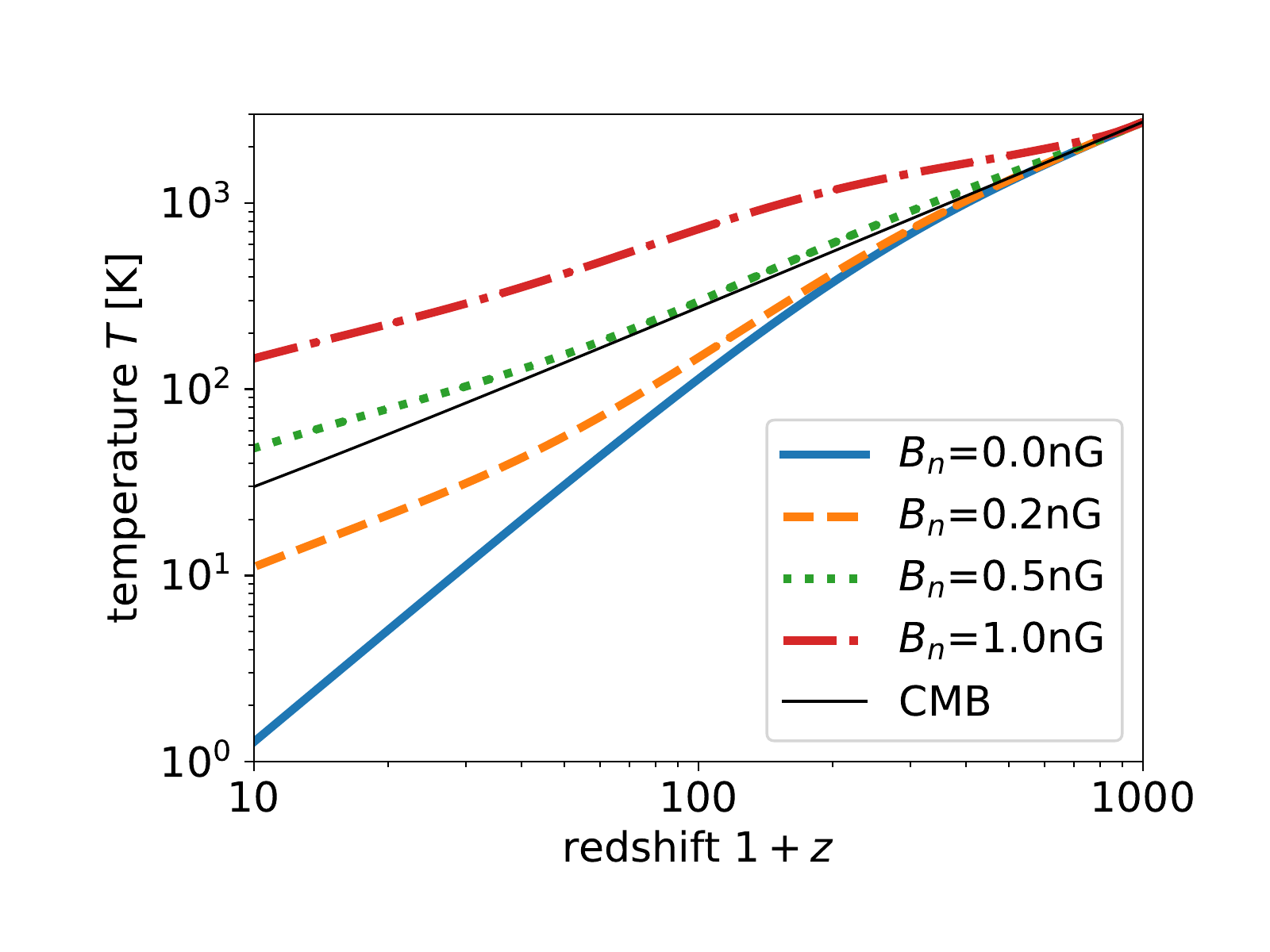}
\caption{
Thermal history of baryon gas including the effects of PMFs.
Different lines represent different strengths of PMFs.
The thick solid line is the case with no PMFs ($B_n=0$ nG),
and the dashed, dotted, dash-dotted lines are
the cases with $B_n=0.2$ nG, $0.5$ nG, and $1.0$ nG.
Here, the scale-dependence of PMFs is fixed on $n_B=-2.9$.
In comparison, we also show the CMB temperature with the thin solid line.
}
\label{fig:Minoda-temperature}
\end{figure}

Putting them all together, figure~\ref{fig:Minoda-temperature} shows the thermal evolution of baryons with PMFs.
In the figure, $B_n$ represents the magnetic field strength smoothed on $\lambda = 1~\rm Mpc$ in equation~(\ref{eq:smoothedB}) with the spectral index $n_B = -2.9$.
As $B_n$ increases, the heating is effective, in particular, in low redshifts. As a result, the deviation from the thermal history without PMFs becomes significant in the redshifts which can be probed by high-redshift 21-cm observations.

\subsubsection{Impact on the density fluctuations}
PMFs can induce the density fluctuations after recombination~\citep{1978ApJ...224..337W, 1996ApJ...468...28K}.
In the MHD approximation, the Lorentz force can induce the velocity fields of baryon fluids. 
Accordingly, the evolution equations of the density fluctuations with the PMFs are described as
\begin{equation}
{\partial^2 \delta_{\rm b} \over \partial t^2} = -2 {\dot a \over
a}{\partial \delta_{\rm b} \over \partial t } +4 \pi G (\rho _{\rm b}
\delta_{\rm b} + \rho _{\rm dm} \delta_{\rm dm} ) + S(t,\boldsymbol{x}),
\label{baryon-den-eq}
\end{equation}
\begin{equation}
S(t,\boldsymbol{x})={ \nabla \cdot \left( (\nabla \times {\boldsymbol{B} (\boldsymbol{x})}) \times {\boldsymbol{B} (\boldsymbol{x})}\right) \over 4 \pi \rho_{{\rm b}, 0}
a^3 (t) },
\end{equation}
\begin{equation}
{\partial^2 \delta_{\rm dm} \over \partial t^2} = -2 {\dot a \over
a}{\partial \delta_{\rm dm} \over \partial t } +4 \pi G (\rho _{\rm b}
\delta_{\rm b} + \rho _{\rm dm} \delta_{\rm dm} ) ,
\label{dm-den-eq}
\end{equation}
where $\rho_{\rm dm}$ is the dark matter density, and $\delta_{\rm b}$ and $\delta_{\rm dm}$ are the density contrasts of baryons and dark matter, respectively.

According to these evolution equations, the density fluctuations of baryons can first grow by the Lorentz force of PMFs, and subsequently, the dark matter density fluctuations catch up with those of baryons due to the gravitational fields described in the last term of equation~(\ref{dm-den-eq}).
In order to solve equations~(\ref{baryon-den-eq})-(\ref{dm-den-eq}), it is useful to introduce the matter density contrast
\begin{equation}
\delta_{\rm m} \equiv { (\rho _{\rm b} \delta_{\rm b} + \rho _{\rm dm} \delta_{\rm dm} ) \over \rho_{\rm m}}, \quad
\rho_{\rm m} \equiv \rho_{\rm b}+\rho_{\rm dm}.
\label{rhom-defi}
\end{equation}
The evolution for $\delta _m $ can be written in
\begin{equation}
{\partial^2 \delta_{\rm m} \over \partial t^2} = -2 {\dot a \over
a}{\partial \delta_{\rm m} \over \partial t } +4 \pi G \rho _{\rm m}
\delta_{\rm m} + {\rho_{\rm b} \over \rho_{\rm m}} S(t,\boldsymbol{x}).
\label{matter-den-eq-2}
\end{equation}

During the matter dominated epoch, the solution of equation~(\ref{matter-den-eq-2}) is given as
\begin{equation}  
\delta_{\rm m} = {\Omega _{\rm b} \over \Omega _{\rm m}}
\left[{9 \over 10} \left({t \over t _{\rm i} }\right)^{2/3} 
+{3 \over 5} \left({t \over t _{\rm i} }\right)^{-1} -{3 \over 2}\right] t_{\rm  i} ^2 S( t_{\rm i}, \boldsymbol{x}),
\label{mag-part}
\end{equation}
with $t_{\rm i}$ being the initial time for generation of the density fluctuations.
As described below, we set $t_{\rm i}$ to the recombination epoch.
Before the recombination epoch, the baryons tightly coupled with photons.
Therefore, the velocity fields of the baryon-photon fluid induced by PMFs cannot grow to the density fluctuations because photon pressure prevents the gravitational growth.
The density fluctuations can evolve after the recombination epoch, when the kinematic coupling of baryons with photon terminates.

Equation~(\ref{mag-part}) tells us the generated density fluctuations grow proportionally to the scale factor, $a \propto t^{2/3}$.
This is because, once the seeds of the density fluctuations are created by PMFs, their evolution is dominated by the gravitational force, implying that the growth rate of the density fluctuations due to PMFs is the same as the one of the primordial density fluctuations generated from the quantum fluctuations during the inflationary expansion. PMF-induced density fluctuations including the non-linear evolution are also estimated by analytically~\citep{2014JCAP...08..017S} and numerically~\citep{2015MNRAS.453.3999M,2016MNRAS.456L..69M}.
Thanks to their estimations, the number count of the collapsed objects or the matter power spectrum will be a good probe of PMFs.

\subsection{Probing the signatures of PMFs with future radio interferometers}
\label{sec:PMFs_obs}

In the previous subsection, we have discussed the cosmological impacts of PMFs: heating the IGM and the generation of the additional density fluctuations.
These impacts affect the evolution of redshift 21-cm signals.
This means that future redshifted 21-cm surveys including SKA could be an important probe of PMFs.
In this subsection, we also review three different topics concerning such centimeter/meter-wave radio observations: the spatially-resolved 21-cm line signal, the 21-cm global signal, and the gravitational wave (GW) observations with the Pulsar Timing Array (PTA).

\subsubsection{spatially-resolved 21-cm line signal}
\cite{2006MNRAS.372.1060T} has firstly 
argued that redshifted 21-cm signal observed by future radio interferometer telescopes can provide a strong constraint on the PMFs.
They have shown that the PMF heating shifts the absorption signal to the emission signal of redshift 21-cm line on the CMB frequency spectrum even in the dark ages. Because the time dependencies of the adiabatic cooling for photons and baryons are different, the gas temperature is always lower than the CMB temperature before the EoR in the case without the heating for baryons including PMFs.
However, when PMFs exist, they can heat up baryons higher than the CMB.
The critical redshift at which the gas temperature surpasses the CMB one depends on the magnetic field strength.
As the magnetic field strength increases, the critical redshift becomes higher.
They have further shown that the small-scale density fluctuations induced by the PMFs enhance the 21-cm fluctuations on small scales.
The enhanced signals are measured as the blue-tilted angular spectrum of 21-cm fluctuations, and the scale-dependence of the 21-cm power spectrum depends on the PMF spectral index $n_{\rm B}$.

PMFs do not only increase the background kinetic gas temperature, but also give fluctuations on the IGM temperature when PMFs are tangled.
\citet{2014PhRvD..89j3522S} has investigated the temperature fluctuations by numerical simulation and studied the effect of baryon temperature fluctuations on the fluctuations of redshifted 21-cm signals. Their results represent that the effect of the temperature fluctuations is negligible and the important source of 21-cm fluctuations due to the PMFs is the small-scale density fluctuations induced by the PMFs.

So far, we have discussed that PMFs can leave a footprint on the future 21-cm line measurements by affecting the IGM density fluctuations and thermal history.
It is worth mentioning that PMFs can change the redshift evolution of the 21-cm signal through the 
non-linear effects on the structure formation history. Thus, we discuss the impact of PMFs on the Jeans length for collapsed object formations.
\citet{2009ApJ...692..236S} has investigated the PMF heating effect on the structure formation and 21-cm signals.
They focused that high baryon temperature heated by the PMFs increases the Jeans scale and delays the star and galaxy formations.
They have pointed out that the delays might decrease the production of Ly-$\alpha$ background and make 21-cm signals small. 
\cite{2009JCAP...11..021S} has also discussed the impact of the magnetic Jeans scale on the non-linear structure formation.
They have found that the PMFs can leave a characteristic feature on the HI two-point correlation function by calculating the ionization bubbles based on a semi-analytic method~\citep{2004ApJ...613....1F}.
Recently, numerical simulations have been conducted to predict a more realistic 21-cm signal with PMF-induced density fluctuations~\citep{2019JCAP...01..033K}.
Still, the fully non-linear simulation including the magnetohydrodynamics, radiative transfer, and the dissipative magnetic fields has not been completed.
Such a comprehensive calculation will be important to improve the constraint on PMFs in the SKA-era.

\subsubsection{21-cm global signal}
\label{sec:21-cm global signal}
\begin{figure}[ht!]
\centering
\includegraphics[width=0.6\textwidth]{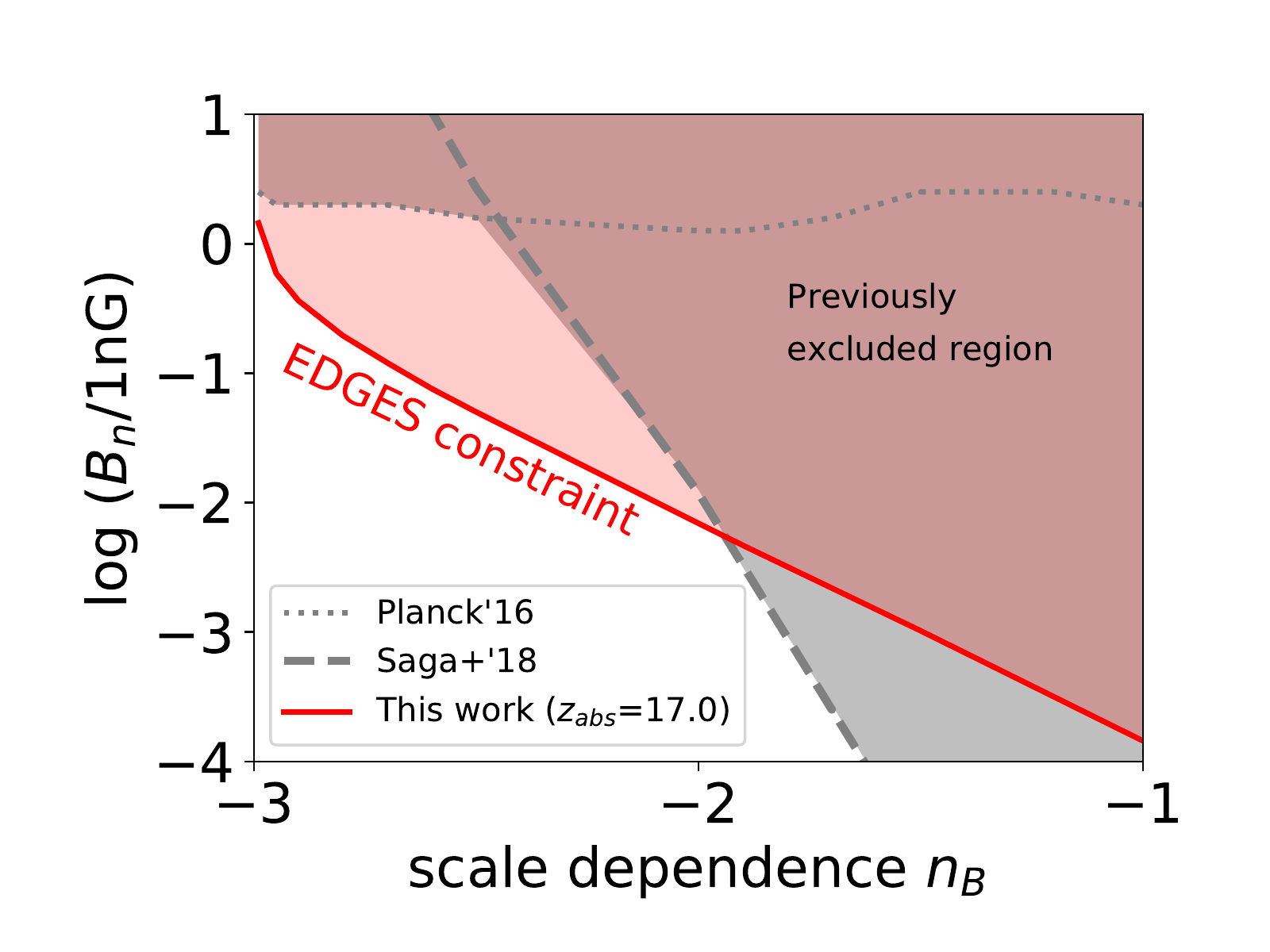}
\caption{
We show a constraint on the PMF strength
depending on the scale dependence with a red thick solid line,
which is obtained from the 21-cm global signal
with the absorption feature at redshift $z=17.0$.
The PMFs with parameters on a red shaded region are ruled out 
because the spin temperature exceeds the radiation temperature.
For comparison, some previous constraints
from CMB anisotropies~\citep{2016A&A...594A..19P}
and magnetic reheating~\citep{2018MNRAS.474L..52S} are also plotted
with gray dotted and dashed lines, respectively.
This figure is referred from figure~2 of~\cite{2019MNRAS.488.2001M}.
}
\label{fig:Minoda}
\end{figure}

In addition to the redshifted 21-cm line fluctuations measured by the future radio interferometers including SKA, the 21-cm global signal, which is the all-sky averaged intensity, can be a useful probe to constrain the PMF strength.
The recent EDGES measurement has reported the detection of the absorption signals in the frequency range corresponding to $15 \lesssim z \lesssim 20$.
This report provides a motivation to obtain the constraint on PMFs from the EDGES measurement results.
\citet{2019MNRAS.488.2001M} has studied the redshift evolution of the global 21-cm signals with PMFs and provided a constraint on the PMF parameters,~$B_n$ and $n_B$.
The important feature of the global 21-cm signal with PMFs is the transition from the absorption to the emission.
When the gas temperature is lower than the CMB temperature, the signal is observed as the absorption on the CMB blackbody spectrum.
On the other hand, when the gas temperature is higher than the CMB temperature, we will observe the 21-cm signal as the emission line.
The EDGES measurement reported that there exists the absorption signal in $15 \lesssim z \lesssim 20$.
Therefore the PMFs which heat up the gas temperature larger than the CMB temperature before $z \sim 15$ are ruled out by the EDGES measurement.
In the parametrization~$B_n$ and $n_B$, the ruled-out region by the EDGES results is shown as the red-colored region in figure~\ref{fig:Minoda}.
For comparison, we plot the constraint by other cosmological observations, Planck CMB observations~\citep{2016A&A...594A..19P} and the magnetic reheating before the recombination epoch~\citep{2018MNRAS.474L..52S}. 
Figure~\ref{fig:Minoda} shows that, for the spectral index, $-3.0 < n_B < -2.0$, the EDGES measurement can provide the tightest constraint on the PMF amplitude.

As mentioned above, the density fluctuations induced by the PMFs might also affect the 21-cm signals. 
The induced density fluctuations can enhance the abundance of small dark matter halos in the early universe, where first stars can form.
First stars can heat and ionize the baryon gas by emitting UV photons.
Therefore, the thermal history of gas is also affected.
\citet{2021MNRAS.507.1254K} has taken into account the enhancement of the first star formation due to the PMF based on their cosmological simulation on the structure formation with PMFs. Accordingly, they updated the constraint on the PMFs from the EDGES measurement. However the modification with including the first star formation enhancement is not relatively strong, compared with the effect of the direct heating due to the PMF energy dissipation mentioned above.

It is worth noting that the EDGES measurement attracts attention to the dark matter model including small coupling with baryons to explain the anomalous signal.
In this context, \citet{Bera:2020jsg} has studied the global 21-cm signals considering both the baryon-dark matter coupling and PMFs.
In this model, the PMF constraint is relaxed more than those shown in figure~\ref{fig:Minoda}, because the baryon-dark matter coupling acts as the cooling source for baryon temperature and compensates the heating effect of PMFs.
Refining the 21-cm line measurements with multiple redshift ranges would be helpful to disentangle several effects from the global signal.

\subsubsection{GW observations with PTA}
So far we have focused on the effect of PMFs on the 21-cm signals.
Actually, SKA, which is one of the future 21-cm observations, has the potential to provide the tighter constraint on PMFs.
On the other hand, SKA can be used as a good GW detector.
Since the PMF can create GWs, the detection or non-detection of GWs by SKA also gives a constraint on the PMFs.
In the following, we provide a short review of the PMF constraint from the GW measurement by SKA.

When PMFs are generated before the neutrino decoupling era, anisotropic stress of PMFs can source GWs both on super-horizon scales and on sub-horizon scales, the so-called passive tensor mode~\citep{2010PhRvD..81d3517S}, as similarly to the passive scalar mode.
Therefore, the direct measurements of GWs offer a unique opportunity to constrain small-scale PMFs.
In this section, following \citet{2001PhRvD..65b3517C,2010PhRvD..81d3517S}, we present the power spectrum of GWs sourced from PMFs.
Using the delta function type of the PMF power spectrum, we show the upper limit on the amplitude of PMFs for various scales based on \citet{2018PhRvD..98h3518S}.
The anisotropic stress in the energy-momentum tensor is a source term of gravitational waves in Einstein equations.
Once PMFs are generated in the early universe, their energy-momentum tensor is given with non-zero anisotropic stress.
Therefore, PMFs can create GWs both on super- and sub-horizon scales after their production~\citep{2001PhRvD..65b3517C,2010PhRvD..81d3517S}.

\begin{figure}
\begin{center}
\includegraphics[width=0.6\textwidth]{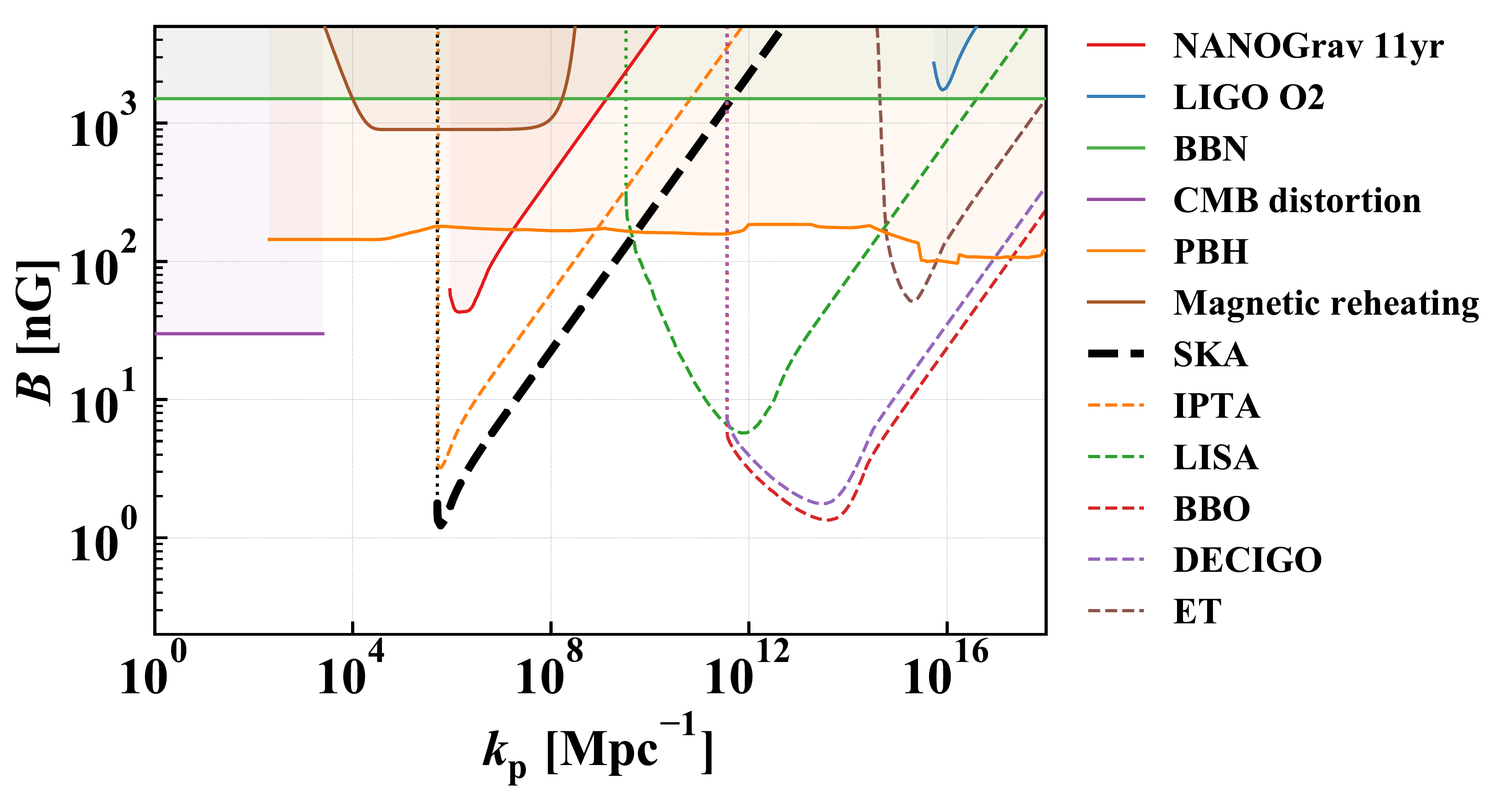}
\end{center}
\caption{
Upper limits on the amplitude of PMFs obtained from direct detection measurements of GWs, fixing $\eta_{B}/\eta_{\nu} = 10^{-12}$ with $\eta_{B}$ and $\eta_{\nu}$ being the conformal time at the PMF generation and neutrino decoupling, respectively.
The shaded region is excluded by the current observations, while the dashed lines indicate the expected upper bound in future experiments.
We also show the upper limit on PMFs by BBN~\citep{2012PhRvD..86f3003K}, magnetic reheating~\citep{2018MNRAS.474L..52S}, CMB distortion~\citep{2000PhRvL..85..700J}, and PBH abundance~\citep{2020JCAP...05..039S}.
({\it updated version of \citet{2020JCAP...05..039S}})
}
\label{fig: limit all}
\end{figure}

In constraining the amplitude of PMFs, we adopt the current upper limit on GWs derived by NANOGrav 11yr results~\citep{2020ApJ...905L..34A} and LIGO O2 results~\citep{2017PhRvL.119p1101A}.
Furthermore, in order to examine the future potential to constrain PMFs, we use the expected sensitivities of the SKA~\citep{2004NewAR..48..979C,2015aska.confE..37J,2020PASA...37....2W}, the International Pulsar Timing Array (IPTA)~\citep{2010CQGra..27h4013H,2013CQGra..30v4010M,2016MNRAS.458.1267V,2018arXiv181010527H}, the Laser Interferometer Space Antenna (LISA)~\citep{2017arXiv170200786A,2019arXiv190706482B}, the Big-Bang Observer (BBO)~\citep{2005PhRvD..72h3005C,2006CQGra..23.2435C}, the Deci-Hertz Interferometer Gravitational-Wave Observatory (DECIGO)~\citep{2001PhRvL..87v1103S,2011PhRvD..83d4011Y,2018PTEP.2018g3E01I}, and the Einstein Telescope (ET)~\citep{2010CQGra..27s4002P,2011CQGra..28i4013H}.
Making use of the current and/or future upper bound on GWs, we present the upper limit on the amplitude of PMFs as a function of the scale $k_{\rm p}$ in figure~\ref{fig: limit all}.
This figure implies that large radio interferometers that probe not only 21-cm signals but also the radio pulsars, such as the SKA, can constrain PMFs through both observations of the 21-cm line and GW observations.

\section{Summary}
In this review, we have brought together the present state of the science case for cosmology using the signals of the 21-cm line emission/absorption during the cosmic dawn and EoR.
We discussed how we can probe the early Universe, particularly focusing on the inflationary era and primordial magnetic fields for illustration.  

As described in section~\ref{sec:inflation}, the primordial power spectrum, more specifically its scale dependence and statistical nature, i.e., primordial non-Gaussianities, can well be measured by future observations of 21-cm fluctuations with SKA and also a futuristic survey. Since 21-cm fluctuations can well probe small scales, by combining observations of CMB on large scales, its scale-dependence which can be described by the so-called running parameters can be well determined. We can also constrain the so-called non-linearity parameters such as $f_{\rm NL}$, $\tau_{\rm NL}$ and $g_{\rm NL}$ by using the 21-cm signals from IGM/minihalo and galaxies. With observations of SKA, we can constrain the non-linearity parameters with the precisions beyond the current sensitivity of CMB, which is almost close to the fundamental limit achievable. Furthermore, as discussed in section~\ref{sec:inflation}, once observations of 21-cm fluctuations during even the dark ages ($z \sim 30-200$) are made available, we can probe non-linearity parameters down to the precision predicted in the standard single-field inflation models, which can give a critical test of the inflationary Universe.
21-cm fluctuations can also probe the adiabaticity of primordial fluctuations. Although isocurvature fluctuations can be severely constrained by CMB on large scales, those on small scales cannot be well limited by CMB. Therefore isocurvature fluctuations which are suppressed on large scales but enhanced on small scales such as the one with the blue-tilted spectrum are still allowed by current observations. However, as mentioned above, 21-cm fluctuations can well probe the small scales, and thus such kind of isocurvature fluctuations can be tested in future observations of 21-cm line. Furthermore, it is also worth noting that 21-cm fluctuations can also differentiate CDM and baryon isocurvature modes, which is impossible by CMB.

In section~\ref{sec:PMFs}, we have discussed the constraint on PMFs from 21-cm line observations. The baryon gas is heated up after the recombination epoch by the dissipative processes of magnetic fields in MHD fluid, which are the ambipolar diffusion and the decaying turbulence. Because not only the kinetic gas temperature, but the spin temperature also arises, the measurements of the 21-cm absorption signal can give a constraint on PMFs. The strength of the heating effect is not homogeneous, and it depends on the distribution of PMFs. In addition, PMFs induce matter density fluctuations due to the Lorentz force. Therefore the 21-cm line fluctuations can be a probe of PMFs. Future observations like SKA will be useful to constrain PMFs.

In the past several decades, precise measurements of the statistical properties of CMB, large-scale structure, supernovae, and so on, 
have established the standard cosmological model and have revealed various aspects of the primordial Universe. However, we still need more precise measurements of 
fundamental observables such as the primordial power spectrum, its adiabaticity, primordial non-Gaussianities, and
primordial magnetic fields, which are necessary to fully understand the history of the Universe.
As shown in this paper, future observations of redshifted 21-cm line emission/absorption signals of neutral hydrogen from cosmic dawn/EoR can probe the above-mentioned observables more precisely, which would give us further insight into our understanding of the primordial Universe.

\begin{ack}

This work was supported in part by JSPS KAKENHI Grant Nos.~17K14304 (D.Y.), 19H01891 (D.Y.), JP20H01932 (S.Y.), JP20K03968 (S.Y.), 17H01131 (T.T.),  19K03874 (T.T.) and 21J00416(S.Y.),  JSPS Overseas Research Fellowships (T.M.), and MEXT KAKENHI Grant No.~19H05110 (T.T.). SY is supported by JSPS Research Fellowships for Young Scientists.

\end{ack}

\bibliographystyle{for_PASJ_sy}
\bibliography{bibforGS.bib} 

\end{document}